\documentclass[12pt]{article}
\usepackage{graphicx}
\usepackage{apacite}

\newcommand{\be}{\begin{equation}}
\newcommand{\ee}{\end{equation}}
\newcommand{\ber}{\begin{eqnarray}}
\newcommand{\eer}{\end{eqnarray}}

\title{Memory capacity of neural network models \thanks{This is a chapter of the forthcoming book "Human memory", Oxford University Press, Edited by M. Kahana and A. Wagner}}
\author{Stefano Fusi \\ \small Center for Theoretical Neuroscience, College of Physicians
	and Surgeons, Columbia University\\ \small Mortimer B. Zuckerman Mind Brain Behavior Institute, Columbia University\\ \small Kavli Institute for Brain Sciences, Columbia University }

\begin{document}
	
	\maketitle
	
\begin{abstract}
Memory is a complex phenomenon that involves several distinct mechanisms. These mechanisms operate at different spatial and temporal levels. This chapter focuses on the theoretical framework and the mathematical models that have been developed to understand how these mechanisms are orchestrated to store, preserve and retrieve a large number of memories. In particular, this chapter reviews the theoretical studies on memory capacity, in which the investigators estimated how the number of storable memories scales with the number of neurons and synapses in the neural circuitry. The memory capacity depends on the complexity of the synapses, the sparseness of the representations, the spatial and temporal correlations between memories and the specific way memories are retrieved. Complexity is important when the synapses can only be modified with a limited precision, as in the case of biological synapses, and sparseness can greatly increase memory capacity and be particularly beneficial when memories are structured (correlated to each other). The theoretical tools discussed by this chapter can be harnessed to identify the important computational principles that underlie memory storage, preservation and retrieval and provide guidance in designing and interpreting memory experiments.
\end{abstract}

\noindent {\it keywords: memory models, memory capacity, memory consolidation, synaptic plasticity, catastrophic forgetting, sparse representations}
\newpage
\tableofcontents

\newpage
\section{Introduction}

Memory is often defined as the mental capacity of retaining information about facts, events, procedures and more generally about any type of previous experience. Memories are remembered as long as they influence our thoughts, feelings, and behavior at the present time. Memory is also one of the fundamental components of learning, our ability to acquire any type of knowledge or skills. 

This chapter focuses on the theoretical framework and the mathematical models of the physical substrate of memory in the biological brain. Basically, any long-lasting alteration of a biochemical process can be considered as a form of memory, although  most of them, if taken individually, cannot really influence our behavior. Identifying those alterations that are important is not easy because memory is a complex  phenomenon that involves several distinct mechanisms. These mechanisms are known to operate at different spatial and temporal scales: some of them involve changes at the level of individual synapses, and some others require complex interactions between entire brain areas. Some of these processes operate on a timescale of milliseconds, while others involve alterations that last a lifetime.

One of the goals of theoretical neuroscience is to try to understand how these processes are orchestrated to store memories rapidly and preserve them over a lifetime. Many theorists have focused on synaptic plasticity, as it is one of the most studied memory mechanisms in experimental neuroscience and it is known to be highly effective when artificial neural networks are trained to perform real world tasks. Some of the synaptic plasticity models are purely phenomenological and they have proved to be important for describing quantitatively the complex and rich observations in experiments on synaptic plasticity. Some other models have been designed to solve computational problems, like pattern classification, or simply to maximize the memory capacity in standard benchmarks. Finally, there are models that are inspired by biology, but then find an application to a computational problem, or vice versa, there are models that solve complex computational problems that then are discovered to be biologically plausible. In this article I will review some of these models and I will try to identify computational principles that underlie memory storage, memory preservation and memory retrieval (see also \cite{fiete16} for a recent review that focuses on similar issues).

The models covered in this chapter are intentionally highly simplified, they often incorporate only a few features of the biological system they imitate, and they are far from being able to capture the complexity and the richness of the phenomenology of human memory. However, they helped us to identify a few general computational principles that might be important for any biological memory system. Many studies focused on the memory capacity of simple neural networks, trying to identify the features of the networks that limit the number of memories that can be stored and then retrieved at a later time. One of the important lessons of these theoretical studies is that some of these limitations might come from the limited precision of the biological elements (e.g. when synaptic weights can be stored only with a few tens of values). The reason why biological synapses are so complex might be related to strategy that overcomes these limitations. Some others constraints are related to the ability to retrieve information. Here we will present the theoretical framework that has been used to formalize some of these memory problems and we will review the main results of the numerous studies on memory capacity.

\section{Memory storage}

\subsection{Storing memories by modifying the weights of synaptic connections}

Artificial neural networks are typically trained by changing the parameters that represent the neuronal activation thresholds and the synaptic weights that connect pairs of neurons. Every time one of these parameters is altered in order to perform a task or simply to acquire information about the sensory world, the parameters are modified and the memory of the neural system is updated. The algorithms used to train artificial neural networks can be divided into three main groups (see e.g. a classic textbook like \citeNP{hkp91} for more details) 1) networks that are able to create representations of the statistics of the world in an
autonomous way (unsupervised learning) 2) networks that can learn to perform a particular task when instructed by a teacher (supervised learning) 3) networks that can learn by a trial and error procedure (reinforcement
learning). These categories can have a different meaning and different nomenclature depending on the community (machine learning or theoretical neuroscience). Again, for all these algorithms, memory is a fundamental component which typically is stored in the parameters of the network. 

\begin{figure}
\centerline{\includegraphics[width=4.5in]{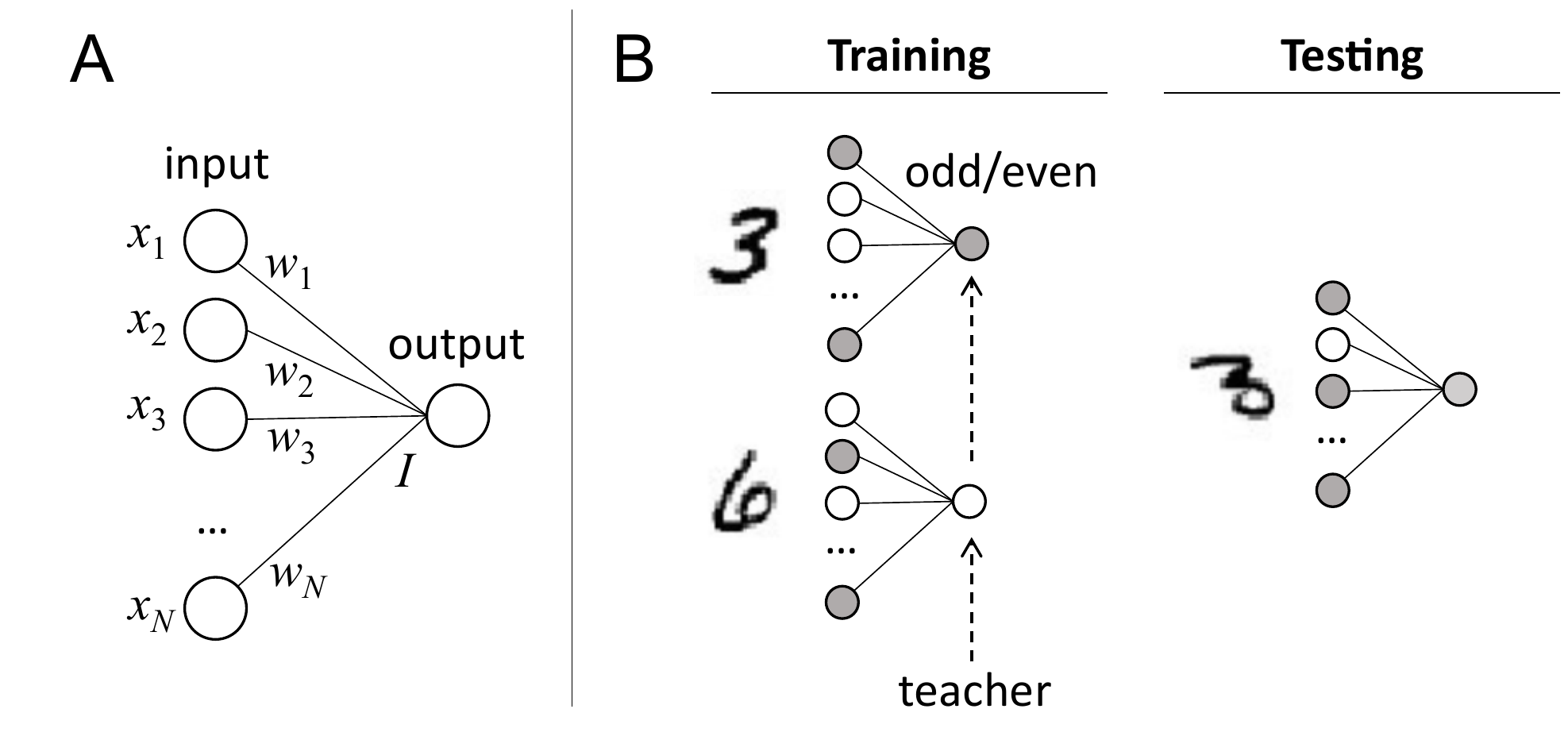}}
	\caption[]{The perceptron in a classification task. A. Schematic of a perceptron: $N$ neurons encode the input. $x_i$ is the state of activation of neuron $i$. These input neurons are connected to the output neuron on the right, which computes the weighted sum of the inputs and compares it to a threshold $\theta$ (not indicated in the figure). B. A simple classification task: the perceptron is trained to report whether a handwritten digit is odd or even. During training one particular sample of the digits is encoded in the input (e.g. an image of a `3', top). Activated neurons are gray. A teacher forces the output neuron to encode the correct response (active for odd digits and inactive for even digits). Following the presentation of each sample, the weights $w$ are updated and when learning terminates they are frozen. During testing a novel digit is presented and the perceptron has to classify it by activating its output if the digit is odd, or remaining silent if the digit is even.}\label{perce}
\end{figure}

\subsubsection{The perceptron}

Rosenblatt \cite{r58, Rosenblatt62} introduced one of the fundamental algorithms for training neural networks. He studied in detail what is probably the simplest feed-forward neural 'network', and the fundamental building block of more complex networks. He called it the perceptron. The perceptron is just one output neuron that is connected to $N$ input neurons. For a given input pattern $x^\mu$ ($x^\mu$ is a vector, and its components $x^\mu_i$s are the activation states of specific neurons), the total current into the output neuron is a weighted sum of the inputs:

$$I^\mu=\sum_{i=1}^N w_{i} x_i^\mu$$

\noindent The output neuron can be either active or inactive. It is activated by the input only when $I$ is above an activation threshold $\theta$. 

The learning algorithm is supervised and it can be used to train the perceptron to classify input patterns into two distinct categories. During learning, the synaptic weights and the activation threshold are tuned so that the output neuron responds to each input as prescribed by the supervisor. For example, consider the classification problem in which the inputs represent images of handwritten digits and the perceptron has to decide whether a digit is odd or even. During training the perceptron is shown a large number of samples of odd and even digits, and the output neuron is set by the supervisor to the activation state corresponding to the class to which the input belongs (e.g. the neuron is activated when the digit is odd, inactivated when it is even). 

The learning procedure ensures that after learning the perceptron responds to an input as prescribed by the supervisor, even in its absence. The input can be one of the samples used for training, or a new sample from a test set. In the second case the perceptron is required to generalize and classify correctly also the new inputs (e.g. a new handwritten digit). 

The proper weights and the threshold of the output neuron are found using an iterative algorithm: for each input pattern, there is a desired output provided by the supervisor, which is $y^\mu$ ($y^\mu=-1$ for input patterns that should inactivate the output neuron and $y^\mu=1$ for input patterns that should activate the output neuron), and each synapse $w_{i}$, connecting input neuron $i$ to the output is updated as follows:

\begin{equation}
w_{i} \to w_{i} + \alpha y^\mu x^\mu_i 
\label{perceptron}
\end{equation}

\noindent where $\alpha$ is a constant that represents the learning rate.
The threshold $\theta$ for the activation of the output neuron is modified in a similar way:

$$\theta \to \theta - \alpha y^\mu$$

The synapses are not modified if the output neuron already responds as desired. In other words, the synapse is updated only if the total synaptic current $I^\mu$ is below the activation threshold $\theta$ when the desired output $y^\mu= + 1$ (and analogously when $I^\mu>\theta$ and $y^\mu=-1$).
To better understand how the learning rule works, consider the case in which $x^1$ is the sample to be learned. Say the response of the perceptron $y=sign(I)=-1$ does not match the one of the teacher, $y^1=+1$. Then the weights are modified so that the vector $w_i$ moves in the direction of the input vector $x_i^1$ and it becomes more aligned to it:
$$w_i \to w_i+\alpha y^1 x^1_i$$
Next time $x^1$ is presented, either for training or for testing, the current $I$ will be larger, closer to or above the threshold $\theta$. Indeed, the new term in the weights will give a positive contribution to the current $I$ that is proportional to the similarity between $x^1$ and the input vector $x$ (the weighted sum essentially computes the scalar product between these two vectors). When $x=x^1$, this contribution will be positive because $y^1=+1$. For other inputs, the contribution might have a different sign, but it will be anyway much smaller provided that the input is sufficiently different from $x^1$. The learning process is then iterated for all input patterns, repeatedly, until all the conditions on the output are satisfied.

The importance of the perceptron algorithm resides in the fact that it can be proved \cite{block1962} to converge if the patterns are linearly separable (i.e. if there exists a $w_{i}$ and a threshold $\theta$ such that $I^\mu>\theta$ for all $\mu$ such that $y^\mu=1$ and $I^\mu<\theta$ for all $\mu$ such that $y^\mu=-1$). In other words, if a solution to the classification problem exists, the algorithm is guaranteed to find one in a finite number of iterations. The convergence proof is probably one of the earliest elegant results of computational neuroscience.

\subsubsection{Hebb's principle}

The perceptron algorithm is also considered one of the early implementations of Hebb's principle \cite{hebb49}. The principle reflects an important intuition of Donald Hebb about a basic mechanism for synaptic plasticity. It states:

{\sl ``When an axon of cell A is near enough to excite a cell B and
	repeatedly or persistently takes part in firing it, some growth
	process or metabolic change takes place in one or both cells such
	that A's efficiency, as one of the cells firing B, is increased.''}

The efficiency he refers to can be interpreted as the synaptic efficacy, or the weight $w_i$ that we defined above. The product of the activities of pre and post-synaptic neurons that appear in the synaptic update equation Eq.\ref{perceptron} is often considered as an expression of the Hebbian principle: when the input and the output neuron (pre and post-synaptic, respectively) are simultaneously active, the synapse is potentiated. In the case of the perceptron, the output neuron is activated by the supervisor during training and it reflects the desired activity. 

\subsubsection{Extensions of the perceptron algorithm}

Besides the perceptron, there are several other learning algorithms that are based on similar principles and often the synaptic weights are modified on the basis of the covariance between the pre and post synaptic activity (see e.g. \cite{sejnowski77,h82}). Many of these algorithms can be derived from first principles, for example by minimizing the error of the output. 

Error minimization is also the basic principle of a broad class of learning algorithms that can train artificial neural networks that are significantly more complex than the perceptron. For example feed-forward networks with multiple layers (deep) can be trained by computing the error at the output and backpropagating it to all the synapses of the network. This algorithm, called backpropagation \cite{rumelhart1986learning}, has recently revolutionalized machine vision, and is extremely popular in artificial intelligence \cite{LeCun2015}. Although it is difficult to imagine how backpropagation can be implemented in a biological system, several groups are working on versions of the algorithm which are more biological plausible (see e.g.\cite{Lillicrap2016, Scellier2016}). More importantly, the neurons of networks trained with backpropagation to perform complex tasks exhibit response properties that are surprisingly similar to those of recorded biological neurons \cite{Yamins2014,Richards2019,Bernardi2020}. For this reason, it is likely that these models will play an important role in understanding the synaptic dynamics and the learning process that lead to the activity observed in the human brain.

\section{Memory preservation}

The synaptic plasticity models described in the previous section determine the desired synaptic modification when a new memory is stored. They have been designed to train artificial networks to perform a particular task and they ignore the complexity of the biological synapses. Deciding the desired synaptic modification is only the first step and it characterizes one of the early phases of biological synaptic plasticity. The consolidation and the maintenance of synaptic modifications require a complex molecular machinery that involves cascades of biochemical processes that operate on different timescales. Typically the process of induction of long term synaptic modifications starts with an alteration of some of the molecules that are locally present at the synapse (e.g. the phosphorylation of CaMKII, see \citeNP{Lisman2002}). 

One of the problems of this type of molecular memory is related to its stability. Because of molecular turnover (see e.g.\cite{crick1984}), the memory molecules are gradually destroyed and replaced by newly synthesized ones. To preserve the stored memories, the state of the old molecules should be copied to the incoming naive ones. If not, the memory lifetime is limited by the lifetime of the molecule, which ranges from a few hours to a few months. None of the molecules that are known to be involved in synaptic plasticity can survive a lifetime. One possible explanation for long memory lifetimes is bistability, which was already proposed by Francis Crick \cite{crick1984}. For example, in the case of CaMKII, an important molecule involved in long term synaptic potentiation, one can imagine that the populations of all molecules has two stable points: one in which none of the molecules is active, and another one in which a large proportion is active. The dynamics of models describing this form of bistability have been studied in detail \cite{lisman1985,lismanzhabotinsky2001,miller2005}.
When a new inactivated protein comes in, it remains unaltered if the majority of the existing CaMKII molecules are inactivated, and it is activated if they are in the active state. As a consequence, the new molecules can acquire the state of the existing ones, preserving the the molecular memory at the level of the population of CaMKII molecules.

CaMKII is only one of the numerous molecules that are involved in synaptic plasticity: more than 1000 different proteins have been identified in the post synaptic proteom of mammalian brain excitatory synapses (see e.g. \cite{EmesGrant2012}). Interestingly, less than 10\% of these proteins are neurotransmitter
receptors, which suggests that the majority of proteins are not directly involved in electrophysiological functions and instead have signaling and regulatory roles. CaMKII is known to be important in the early phases of the induction of long term synaptic potentiation (E-LTP). The molecules involved in E-LTP activate a cascade of biochemical processes that eventually regulate gene transcription and protein synthesis, leading to permanent changes in the morphology of the synaptic connections or to persistent molecular mechanisms that are known to underlie late long term potentiation (L-LTP) maintenance. The foundational work on these cascades of biochemical processes of Eric Kandel and colleagues is summarized in \cite{squirekandel}.

To understand the computational role of these highly organized protein networks, it is necessary to review more than 30 years of theoretical studies. The next sections summarize some of the important results of these studies that show that biological complexity plays a fundamental role in maximizing memory capacity.

\subsection{Memory models and synaptic plasticity}

For many years, research on the synaptic basis of memory focused on the long-term potentiation of synapses which, at least by the modeling community, was represented as a simple switch-like change in synaptic state. Memory models studied in the 1980’s (i.e. \cite{h82}) suggested that networks of neurons connected by such switch-like synapses could maintain a huge number of memories virtually indefinitely. Although it becomes progressively more difficult to retrieve memories in these models as time passes and additional memories are stored (see also the last section), the memory traces of old experiences never fade away completely (see e.g. \cite{a89,hkp91}). Memory capacity, which was computed to be proportional to network size, was only limited by interference from multiple stored memories, which can hamper memory retrieval. This work made it appear that extensive memory performance could arise from a relatively simple mechanism of synaptic plasticity. However, it was already clear from the experimental works summarized above that synaptic plasticity is anything but simple. If, as the theoretical work suggested, this complexity is not needed for memory storage, what is it there for? 

\begin{figure}
\centerline{\includegraphics[width=4.5in]{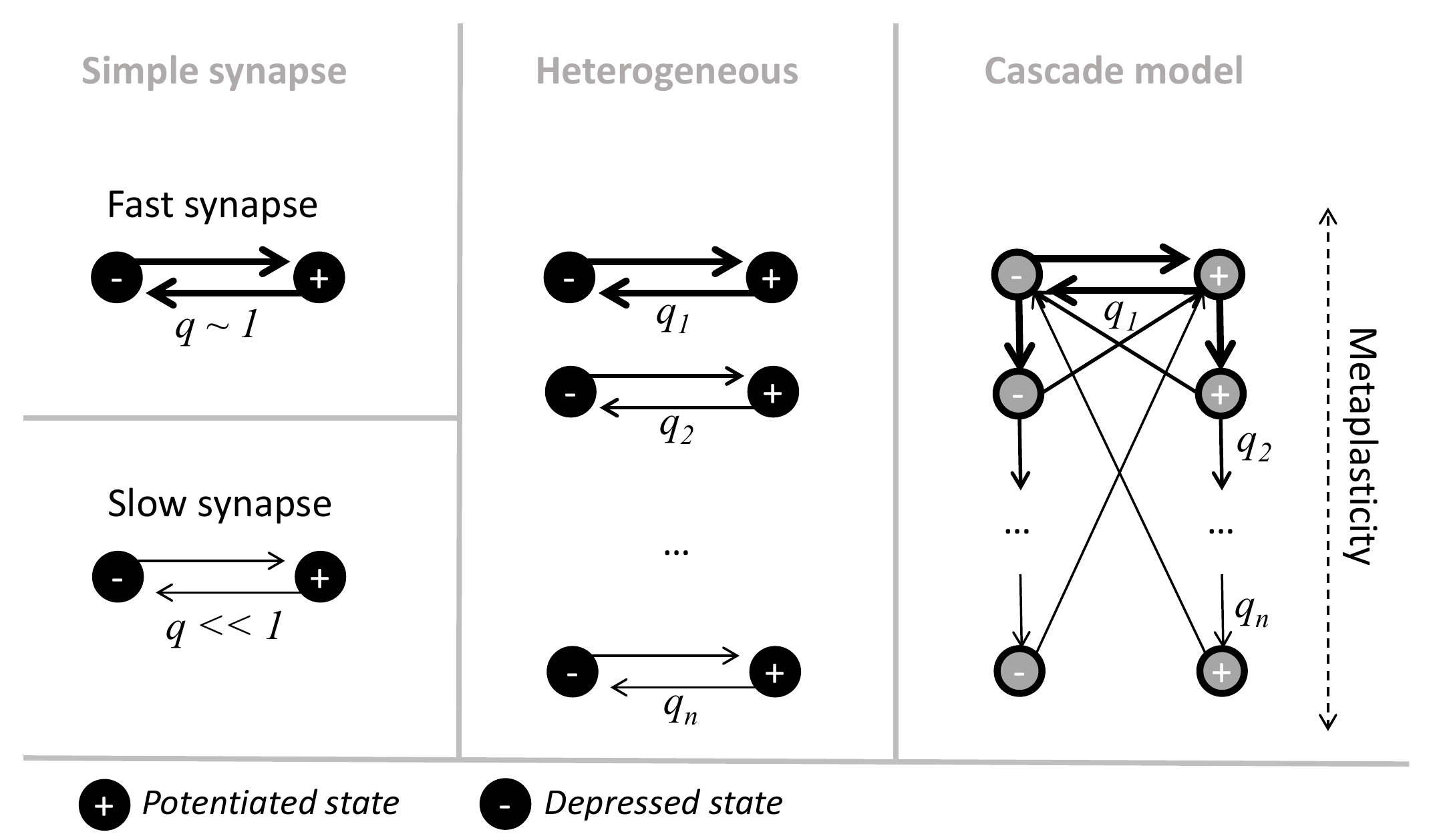}}
	\caption[]{Scheme of three synaptic models. Simple synapse: each synapse has only two states, a potentiated one, indicated with a filled black circle with a '+' and a depressed one, with a '-' inside the circle. Every time a memory needs to be stored by potentiating the synapse, a transition to the potentiated state occurs with probability $q$ (the arrow to the right). Analogously, when the activity imposed on the pre and post-synaptic activity requires a depression, the synapse makes a transition to the depressed state with probability $q$. So $q$ is basically the learning rate: when $q$ is close to 1, the synapse is modified rapidly (fast synapse) and when it is small the synapse is slow. The thickness of the arrow is proportional to the value of $q$. In the heterogeneous model different synapses are characterized by different learning rates $q_k$, so the network has both fast and slow synapses. In the cascade model, each synapse has a series of depressed and a series of potentiated states (gray circles), each characterized by a different learning rate. So each synapse can operate on a different timescale depending on the internal state. Vertical transitions correspond to metaplastic processes: they do not change the synaptic efficacy (potentiated or depressed) but only the internal state of the synapse and hence the $q$ of future plastic changes.}\label{models}
\end{figure}

The key to answering this question arose from work done at the beginning of the 90’s. This work arose from a project led by D. Amit aimed at implementing an associative neural network in an electronic chip using the physics of transistors to emulate neurons and synapses, as originally proposed by Carver Mead \cite{mead}. The main problem encountered in this project was related to the limited memory of the electronic system. The problem was not how to preserve the states of synapses over long times, but how to prevent memories from being overwritten by other memories. Memories were overwritten by other memories so rapidly that it was practically impossible for the neural network to store any information. Subsequent theoretical analysis of this problem \cite{af92,af94,f02,fa07} showed that what had appeared to be a simple approximation made in the theoretical calculations of the 80s was actually a fatal flaw. The unfortunate approximation was ignoring the limits on synaptic strength imposed on any real physical or biological device. As we will explain in detail in the next Sections, when these limits are included, the memory capacity grows only logarithmically rather than linearly with network size. This is a dramatic reduction in performance, especially when one considers that the number of synapses is typically huge (in the human brain it is of the order of $10^{15}$). A model with a logarithmic capacity would be extremely inefficient and several orders of magnitude worse than a model with linear capacity.

\begin{figure}
\centerline{\includegraphics[width=3.5in]{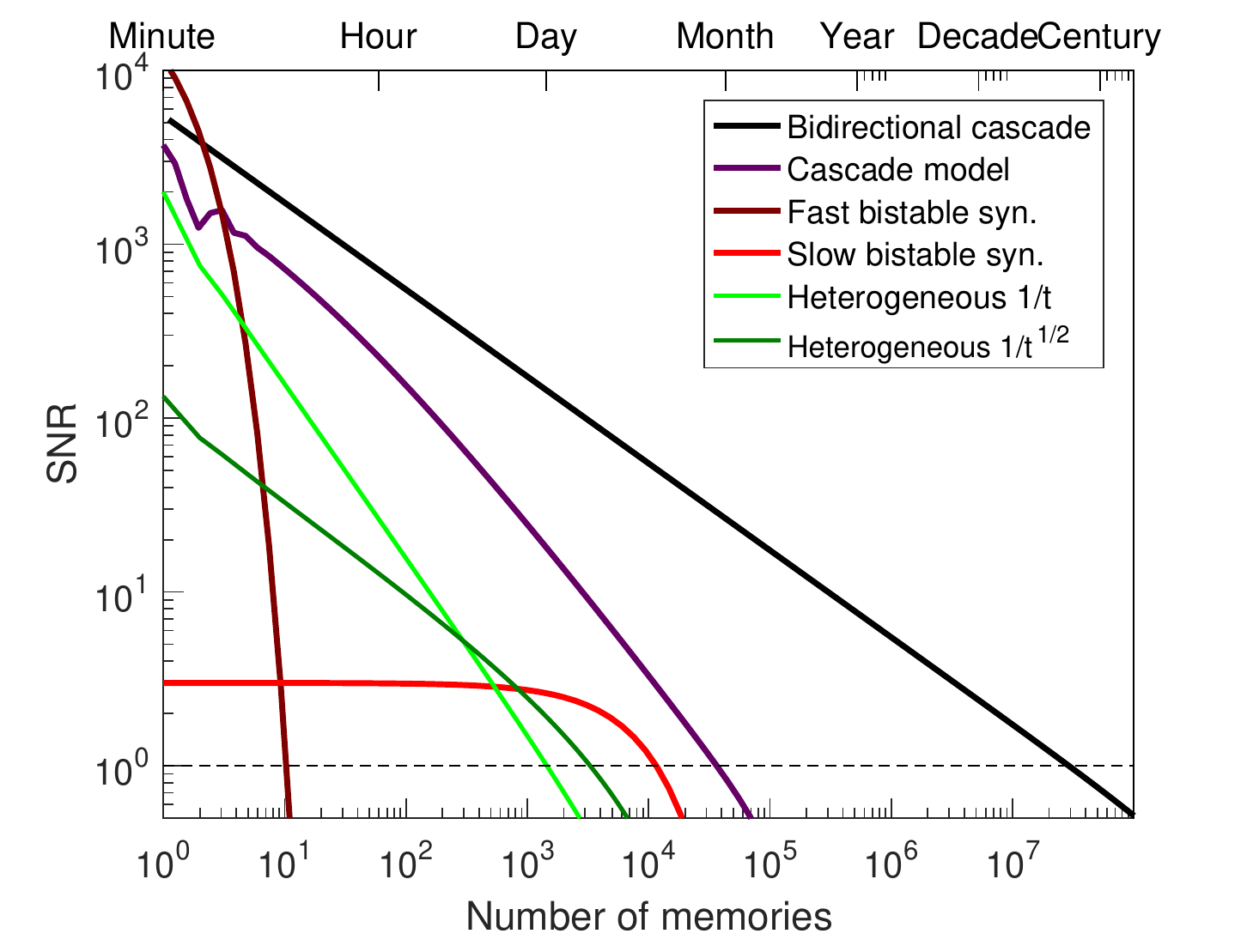}}
	\caption[]{Memory trace (signal to noise ratio, or SNR) as a function of time (i.e. the age of the tracked memory) for four different models. The dashed line is an arbitrary threshold for memory retrieval: memories are basically forgotten when SNR drops below this threshold. 
	Memories decay because other new memories overwrite them, not because time is passing.
	To give an idea of the timescales that might be involved, we indicated at the top the time intervals that correspond to the number of stored memories shown at the bottom assuming that memories are stored at the arbitrary rate of one every minute.
	Dark red line: fast simple bistable synapses (all synapses have $q$ near 1). The initial memory trace is large, but the decay is rapid. Light red line: slow simple bistable synapses (all synapses have small $q \sim 1/\sqrt{N}$): long memory lifetime but small initial memory trace. Purple: cascade model, with a large initial memory trace, power law decay ($1/t$), and long memory lifetimes. Black: bidirectional cascade model: power law decay ($1/\sqrt{t}$) and the initial memory trace is as large as for the cascade model. 
	Light and dark green: heterogeneous population of simple bistable synapses. Light green: synapses are divided in 20 equal size subpopulations, each characterized by a different value of the learning rate $q$ ($q=0.6^{(k-1)}, k=1,...,20$). The SNR decays as $1/t$ and the scaling properties are the same as for the cascade model. Dark green: same number of subpopulations, but now their size increases as $q$ becomes smaller (the size is proportional to $1/\sqrt{q}$). The decay is slower ($1/\sqrt{t}$) compared to the heterogeneous model with equal size subpopulations, however the initial SNR is strongly reduced as it scales as $N^{1/4}$. For this model the memory lifetime scales as $\sqrt{N}$. 
}\label{snr}
\end{figure}

\subsection{The plasticity-stability tradeoff} 

In discussing the capacity limitations of any memory model, it is important to appreciate a tradeoff between two desirable properties, plasticity and stability \cite{cg91}. To reflect this tradeoff, it is useful to characterize memory performance by two quantities \cite{fda05,fa07,bf16,arbib16}. One is the strength of the memory trace right after a memory is stored. This quantity reflects the degree of plasticity in the system, that is its ability to store new memories. The other quantity is memory lifetime, which reflects the stability of the system for storing memories over long times.

To better understand the trade-off it is important to define more precisely what we mean by memory strength. In the standard memory benchmark, the strength of a particular memory trace is estimated in a particular situation in which memories are assumed to be random and uncorrelated. One of the main reasons behind this assumption is that it allowed theorists to perform analytic calculations. However, it is a reasonable assumption even when more complex memories are considered. Indeed, storage of new memories is likely to exploit similarities with previously stored information (consider e.g.~semantic memories). Hence the information contained in a memory is likely to be pre-processed, so that only its components that are not correlated with previously stored memories are actually stored (see also the last section in this Chapter). In other words, it is more efficient to store only the information that is not already present in our memory. As a consequence, it is not unreasonable to consider memories that are unstructured (random) and do not have any correlations with previously stored information (uncorrelated). 

\subsubsection{Memory traces: signal and noise} Consider now an ensemble of $N$ synapses which is exposed to an ongoing stream of modifications which lead to the storage of new memories. For simplicity, we assume that each memory corresponds to a distinct pattern of random and uncorrelated synaptic modifications.  One can then select arbitrarily one of these memories and track it over time. The selected memory is not different or special in any way, so that the results for this particular memory apply equally to all the memories being stored.

To track the selected memory one can take the point of view of an ideal observer that knows the strengths of all the synapses relevant to a particular memory trace \cite{f02,fda05}. In the brain synapses are not read directly, the readout is implemented by complex neural circuitry. So the strength of the memory trace based on the ideal observer approach may be significantly larger than the memory trace that is actually usable by the neural circuits. As a consequence, the memory capacity estimated using the ideal observer can be considered as an upper bound on the actual memory capacity. However, as we will show, there are many situations in which the ideal observer approach predicts the correct scaling properties of the memory capacity of simple neural circuits that actually perform memory retrieval (see the Memory retrieval section).
 
More formally we define the memory signal of a particular memory that was stored at time $t^\mu$ as the overlap (or similarity) between the pattern of synaptic modifications $\Delta w_i$ imposed by the event and the current state of the synaptic weights $w_i$ at time $t$:
\[ \label{DefSignalMT}
{\mathcal{S}}^{\mu}(t) \equiv {1 \over N}  \mathbf{E}\left[  \sum_{i=1}^N w_{i}(t)\,  \Delta w_{i}(t^\mu) \right] \ .
\]
$\mathbf{E}$ indicates an average (expectation) over the random uncorrelated patterns that represent the other memories and that make the trace of the tracked memory noisy. The noise is just the standard deviation of the overlap that defines the signal:
\[ \label{DefNoiseMT}
{\cal N}^{\mu}(t) \equiv \sqrt{ {1 \over N^2}  \mathbf{E}\left[  \Big{(} \sum_{i=1}^N w_{i}(t)\,  \Delta w_{i}(t^\mu) \Big{)}^2 \right] - {\cal S}^{\mu}(t)^2 } \ .
\]
The quantity gives the strength of the trace of memory $\mu$ will then be ${\cal S}/{\cal N}$, the signal to noise ratio (SNR) of a memory. 

\subsubsection{The initial signal to noise ratio: plasticity}

The initial SNR is then the SNR of a memory immediately after it has been stored, when it is most vivid. Highly plastic synapses allow for large initial SNR. In all realistic models, the SNR decreases with the memory age, either because old memories are overwritten by new ones, causing a degradation of the signal, or because the interference of new memories increases the noise (see the next sections for example models of these scenarios). So the initial SNR is often the largest SNR, especially under the assumption that the memories to be stored are dissimilar to each other (random and uncorrelated). It is desirable to have a large SNR (and hence a large initial SNR) because the SNR is related to the ability to retrieve a memory from a potentially noisy cue (see e.g.\cite{h82,af94,bf16}). Typically there is a threshold above which a memory becomes retrievable. This threshold depends on the architecture and the dynamics of the neural circuits that store the memory, but also on the nature of the cue that triggers memory retrieval. Highly effective cues can retrieve easily the right memory, whereas weak retrieval cues might lead to the recall of the wrong memory. In the case of random uncorrelated memories it is possible to define more precisely what an effective cue is. For example, it is possible to train a perceptron to classify random input patterns and then retrieve memories by imposing on the input neurons degraded versions of the stored patterns. Degraded inputs can be obtained, for example, by changing randomly the activation state of a certain fraction of input neurons. The input patterns that are most similar to those used during training and hence stored in memory are the most effective retrieval cues. They are more likely to be classified correctly than highly degraded inputs. Higher SNR means a better ability to tolerate degradation. More quantitatively, the minimum overlap between the input and the memory to be retrieved that can be tolerated (i.e.~that produces the same response as the stored memory) is inversely proportional to the SNR\cite{Krauth1988,bf16}. This dependence demonstrates the importance of large SNRs: classifiers whose memory SNR is just above retrieval threshold can correctly recognize the inputs that have been used for training, but they will not necessarily generalize to degraded inputs. For generalization higher SNRs are needed.

\subsubsection{Memory lifetime and stability} 

Now that we have introduced a quantity that reasonably represents memory strength, we can also define more precisely the memory lifetime as the maximal time since storage over which a memory can be detected, i.e.~for which the SNR is larger than some threshold. Stable memories have long memory lifetime. The SNR threshold, as discussed above, depends on the details of the neural circuit and on the nature of the stored memories. However, the scaling properties of the memory performance do not depend on the precise value of the threshold. If new memories arrive at a constant rate, the lifetime is proportional to the memory capacity, because memories that have been stored more recently than the tracked one will have a larger SNR, and hence if the tracked memory is likely to be retrievable, so are more recent ones. 
The actual memory capacity of neural networks will depend on many details and in particular on the neural dynamics, as we will discuss extensively in section \ref{memretrieval}. 

\subsubsection{Unbounded synapses}

In the case of the models of the 80's, like the Hopfield model \cite{h82}, the memory signal is constant over time, despite the storage of new uncorrelated memories. Indeed, every time a new memory is stored, the synapses are modified according to a simple Hebbian rule:
$$ w_{ij} \to w_{ij} + \alpha x_i x_j$$
where $i$ is the index of the pre-synaptic neuron, $j$ the index of the post-synaptic neurons, $x_i$ is the state of activation of neuron $i$ when the pattern of activity representing the memory to be stored is imposed to the network and $\alpha$ is a constant that characterizes the learning rate. As new memories are added linearly, all memories are equivalent and there is no mechanism that favors new over old memories (but see \cite{Kahana2012}, Chapter 5, for an extension of the Hopfield model in which it is possible to observe a recency effect, i.e. a greater weight to recent memories). The SNR decreases and memories become irretrievable only because the memory noise becomes too large due to the interference between memories. In the case of random memories the noise increases as $\sqrt{t}$ and hence it depends only on the total number of memories stored until time $t$. The SNR also increases with the size of the network. More specifically it is proportional to $\sqrt{N}$, where $N$ is the number of independent synapses. This means that the SNR crosses the retrieval threshold at a time $t$ that is proportional to $N$, which is a long memory lifetime if one considers that $N$ can be very large in biological brains. This huge memory capacity is due to peculiar dependence of the memory signal on the number of stored memories: as new memories are stored, the signal always remains constant. This peculiarity comes from the assumption that the synaptic weights can grow unboundedly over time, which is clearly unrealistic for any biological system. Rescaling the weights would require synapses that can be modified with a precision that increases over time, which is also unrealistic. When reasonable bounds or limits on the precision are imposed (biological synapses are estimated to have no more than 26 distinguishable states \cite{BartolJr2015}), then the situation is very different, and the memory signal decays very rapidly with time, as discussed in the next section.

\subsubsection{Bounded synapses} 

Consider a switch-like simple synapse whose weight has only two values (i.e. the synapse is bistable as it can be either potentiated or depressed, see Figure \ref{models}). This might sound like a pathological case, but it is actually representative of what happens in a large class of realistic synaptic models (see below). Suppose that a particular pattern of pre- and postsynaptic activity modifies a synapse if it is repeated over a sufficient number of trials.  The parameter $q$, which we use to characterize how labile a synapse is to change, is the probability that this pattern of activity produces a change in a synapse on any single trial. Because synapses with large $q$ values change rapidly, we call them fast, and likewise synapses with small $q$ are termed slow.  This maps a range of $q$ values to a range of synaptic timescales. For a population of synapses with a particular value of $q$, the strength of the memory trace (i.e. the SNR) at the time of storage is proportional to $q$. The memory signal decays exponentially with time, with a time constant that is proportional to $1/q$.
Indeed, for synapses with only two values, every time a synapse is modified, all the information previously acquired and stored in that synapse is lost. Say for example that a synapse is potentiated. Whether it started from the depressed or the potentiated state it will end up in the potentiated state, erasing all the information previously stored and contained in the initial state. We now consider one memory we intend to track: the number of synapses that encode that particular tracked memory will be initially proportional to $q$ (all the synapses actually modified). Then each of these synapses will preserve information about the tracked memory as long as it is not modified by other memories. The probability the a synapse is not modified by each memory is $1-q$, and hence after $t$ memories the probability will be $(1-q)^t$. So the number of synapses that are still encoding the tracked memory will be proportional to:

$$q(1-q)^t = q \exp[\log(1-q)^t] = q \exp[t\log(1-q)] \simeq q \exp(-qt)$$

\noindent where we used the definition of the exponential and the logarithm in the first step and we Taylor expanded the logarithm in the last step, assuming that $q$ is small enough. This expression clearly shows that the memory lifetime goes as $1/q$. This inverse dependence is a mathematical indication of the plasticity-stability tradeoff. 

In non-mathematical terms, synapses that are highly labile quickly create memory traces that are vivid right after they are stored but that fade rapidly (Fig.~\ref{snr} - fast bistable synapses). Synapses that resist change and are therefore slow are good at retaining old memories, but bad at representing new ones (Fig.~\ref{snr} - slow bistable synapses).  In Fig.~\ref{snr} we plotted the memory SNR in these two cases. Notice that the horizontal and vertical scales in the figure are both logarithmic so all the differences seen are large. For example, fast synapses have an initial memory strength that is orders of magnitude larger than slow synapses. For fast synapses it is proportional to $\sqrt{N}$, where $N$ is the number of independent synapses, whereas for slow synapses it does not scale at all with $N$. However, the memory lifetime is orders of magnitude smaller for fast synapses (it scales as $\log N$, compared to the $\sqrt{N}$ scaling of slow synapses). 

Here we discussed the case of bistable synapses, but the plasticity stability trade-off is very general and it basically applies to any reasonably realistic synaptic model. For example, for synapses that have to traverse $m$ states before they reach the bounds, the memory capacity increases at most by a factor $m^2$, but it is still logarithmic in $N$ \cite{fa07}. The logarithmic dependence is preserved also when soft bounds are considered \cite{fa07} (see also \cite{van2012soft} for an interesting comparison between hard and soft bound synapses).  Given the generality of the plasticity-stability trade-off, how can we rapidly memorize so many details about new experiences and then remember them for years?

\subsection{The cascade model of synaptic plasticity: the importance of complexity}

The solution proposed in \cite{fda05} is based on the idea that if we want the desirable features of both the fast and the slow synapses, we need synaptic dynamics that operate on both fast and slow timescales. Inspired by the range of molecular and cellular mechanism operating at the synaptic level, in the model proposed in \cite{fda05}, called the "cascade model",  $q$ depended on the history of synaptic modifications. The synapse starts in either a potentiated or a depressed state. These two states, shown at the top of the diagram in Figure \ref{models} are the most plastic and are characterized by a learning rate $q$ that is close to 1, as in the fast synapses discussed earlier. When the synapse is in the potentiated state and it needs to be further potentiated, it makes a transition to a 'hidden' state that corresponds to the same synaptic efficacy, but is more resistant to depression. Indeed, the synapse can be depressed only with a probability $q_2<q_1$. If the synapse needs to be depressed and it actually makes a transition to the depressed state, then it is reset to the most plastic state at the top of the cascade. The diagram is completed by a full set of progressively more 'rigid' hidden states ($q_1>q_2>...>q_n$), both for the potentiated and the depressed state. In \cite{fda05} the values of the $q$s decrease exponentially ($q_k=Q^k$ where $Q$ is a constant), so that they can cover timescales that vary over multiple orders of magnitude. There is evidence that biological synapses become more resistant to depression after a long series of potentiations \cite{oconnor2005} and that learning occurs on multiple timescales that range from hundreds of milliseconds to months \cite{Fusi2007,iigaya2019}.

Although all the synapses in this model are described by the same equations, at any given time their properties are heterogeneous because their different histories puts them in different states that correspond to different values of the learning rate $q$. This history dependence is call metaplasticity \cite{abraham1996}, or plasticity of plasticity.
This improves the performance of the model dramatically and it suggests why synaptic plasticity is such a complex and multi-faceted phenomenon. The cascade model is characterized by a memory signal that decays as $1/t$. Both the initial SNR and the maximum memory lifetime scale as $\sqrt{N}$, where $N$ is the number of synapses. Interestingly, there is evidence that some human forgetting processes are well described by power laws \cite{wixted1997}. The observed power laws typically have a slower decay than those derived from the models, but this is expected given that the memories considered in the experiments are not random and uncorrelated (see the section below about correlated memories).

The cascade model is an example of a complex synapse that does significantly better than simple synapses. However, its scaling properties are not different from those of a heterogeneous population of simple synapses (see Figure \ref{models}) in which different synapses are characterized by different values of $q$ \cite{fda05,rf13} (Fig \ref{snr}, see heterogenous models with $1/t$ decay). The interactions between fast and slow components increase significantly the numerical value of the SNR, but not its scaling properties. It is only with the recent bidirectional cascade model described below that one can improve scalability.

\subsection{The bidirectional cascade model of synaptic plasticity: complexity is even more important} 

\begin{figure}
	\centerline{\includegraphics[width=4.5in]{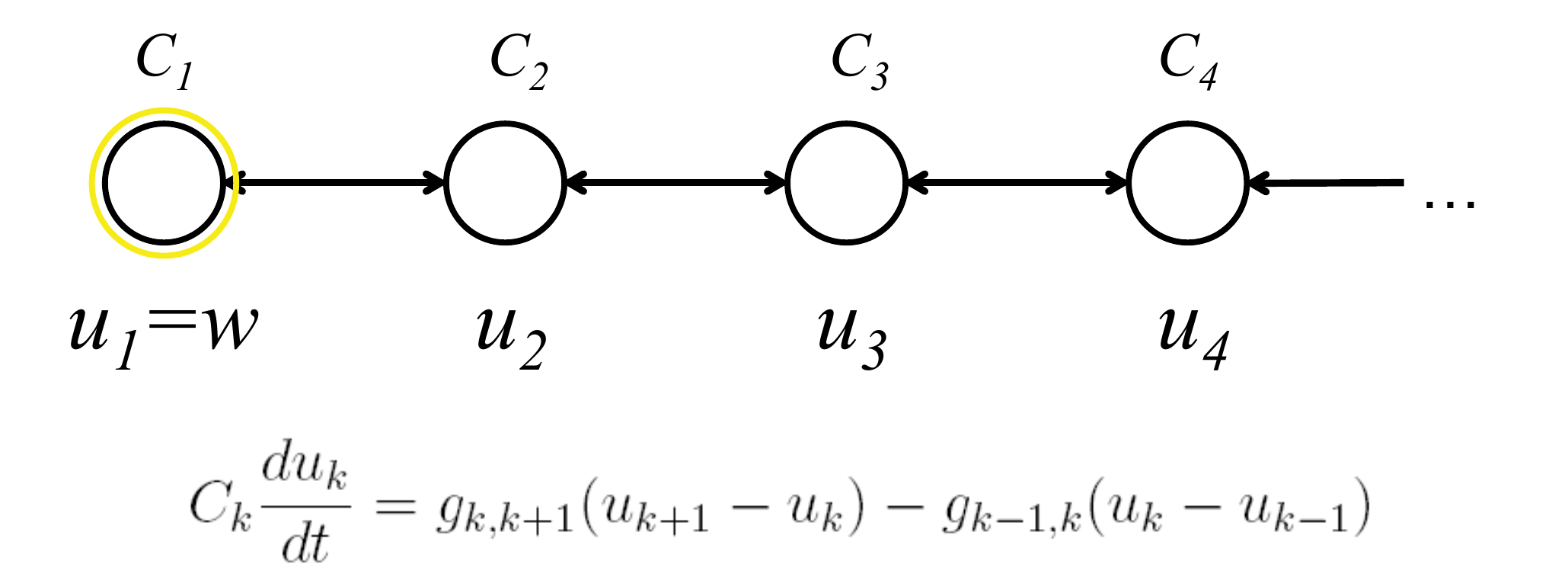}}
	\caption[]{\small The bidirectional cascade model of a single synapse: The dynamical variables $u_k$ represent different biochemical processes that are responsible for memory consolidation ($k=1,...,m$, where $m$ is the total number of processes). They are arranged in a linear chain and interact only with their two nearest neighbors (see differential equation), except for the first and the last variable. The first one interacts only with the second one (and is also coupled to the input), while the last one interacts only with the penultimate one. Moreover, the last variable $u_m$ has a leakage term that is proportional to its value (obtained by setting $u_{m+1}=0$). The parameters $g_{k,k+1}$ are the strengths of the bidirectional interactions (double arrows). Together with the parameters $C_k$ they determine the timescales on which each process operates. The first variable $u_1$ represents the strength of the synaptic weight.
	}
\label{bf}
\end{figure}

Bidirectional cascade models are actually a class of functionally equivalent models that are described in \cite{bf16}. Fig. \ref{bf} shows one possible implementation, a simple chain model that is characterized by multiple dynamical variables, each representing a different biochemical process. The first variable, which is the most plastic one, represents the strength of the synaptic weight. It is rapidly modified every time the conditions for synaptic potentiation or depression are met. Unlike the cascade model explained in the previous section, this variable is not bistable, but it has multiple values that correspond to different synaptic strengths. The other dynamical variables are hidden (i.e. not directly coupled to neural activity) and represent other biochemical processes that are affected by changes in the first variable. In the simplest configuration, these variables are arranged in a linear chain, and each variable interacts with its two nearest neighbors. These hidden variables tend to equilibrate around the weighted average of the neighboring variables. When the first variable is modified, the second variable tends to follow it. In this way a potentiation/depression is propagated downstream, through the chain of all variables. Importantly, the downstream variables also affect the upstream variables as the interactions are bidirectional. This makes the synapse metaplastic, as the cascade synaptic model described above. The dynamics of different variables are characterized by different timescales, which are determined in the simple example of Fig. \ref{bf} by the $g$ and $C$ parameters. More specifically, the variables at the left end of the chain are the fastest, and the others are progressively slower. When the parameters are properly tuned, the initial SNR scales as $\sqrt{N}$, as in the cascade model previously discussed, but the memory lifetime scales as $N$, which, in a large neural system, is a huge improvement over the $\sqrt{N}$ scaling of previous models. The memory decay is approximately $1/\sqrt{t}$, as shown in Fig. \ref{snr}. The model requires a number of dynamical variables that grows only logarithmically with $N$ and it is robust to discretization and to many forms of parameter perturbations. The model is significantly less robust to biases in the input statistics. When the synaptic modifications are imbalanced the decay remains almost unaltered, but the SNR curves are shifted downwards. The memory system is clearly sensitive to imbalances in the effective rates of potentiation and depression.

In the bidirectional cascade model the interactions between fast and slow variables are significantly more important than in previous models. Indeed, it is possible to build a system with non-interacting variables that exhibits a $1/\sqrt{t}$ decay. However, this requires disproportionately large populations of slow variables, which greatly reduce the initial SNR (it scales only as $N^{1/4}$). For these heterogeneous models the memory lifetimes scales only like $\sqrt{N}$.

\subsection{Biological interpretations of computational models of complex synapses}

One possible interpretation of the dynamical variables $u_k$ is that they represent the deviations from equilibrium of chemical concentrations. The timescales on which these variables change would then be determined by the equilibrium rates (and concentrations) of reversible chemical reactions. However, for the slowest variables, which vary on timescales of the order of years, it is probably necessary to consider biological implementations in which the $u_k$ correspond to multistable processes. For example, the slowest variable could be discretized, sometimes with only two levels \cite{bf16}, and hence they could be implemented by a bistable process, which would allow for very long timescales \cite{crick1984,miller2005}. For a small number of levels that is larger than two, one could combine multiple bistable processes or use slightly more complicated mechanisms \cite{Shouval2005}. These biochemical processes could be localized in individual synapses, and recent phenomenological models indicate that at least three such variables are needed to describe experimental findings \cite{Ziegler2015}.

However, these processes could also be distributed across different neurons in the same local circuit or even across multiple brain areas. The interaction between two coupled $u_k$ variables could be mediated by neuronal activity, such as the widely observed replay activity (see e.g.\cite{rf13}). In the case of different brain areas, the synapses containing the fastest variables might be in the medial temporal lobe, e.g.~in the hippocampus, and the synapses with the slowest variables could reside in the long-range connections in the cortex. This is an important and old idea already proposed in \cite{mcclelland95} and recently discussed in \cite{KUMARAN2016} (see also the section about Temporal correlations and catastrophic forgetting below).

Several experimental studies on long term synaptic modifications have revealed that synaptic consolidation is not a unitary phenomenon, but consists of multiple phases. One particularly relevant example is related to studies on hippocampal plasticity and more specifically to what is known as the synaptic tagging and capture (STC) hypothesis, which explains several experimental observations. According to the STC hypothesis, LTP consists of at least four steps \cite{Reymann2007, Redondo2011}: first, the expression of synaptic potentiation with the setting of a local synaptic tag; second, the synthesis and distribution of plasticity related proteins (PrPs); third, the capture of these proteins by tagged synapses; and forth, the final stabilization of synaptic strength. Phenomenological models \cite{clopath08, Barrett2009, Ziegler2015} of STC comprise all four steps, and can explain experiments on the induction of protein synthesis dependent late LTP. The model dynamics of \cite{clopath08,Barrett2009} are characterized by four dynamical variables: the first two are tag variables, one for LTP and one for LTD. They could correspond to two variables that are modified to induce LTP and LTD. The authors of \cite{clopath08} hypothesized that a candidate molecule involved in the tag signaling could be CaMKII. The third variable describes the process that triggers the synthesis of PrPs and the fourth one the stabilization of the synaptic modification. A candidate protein involved in the maintenance of potentiated hippocampal synapses is the protein kinase M$\zeta$ (PKM$\zeta$). The PrPs that are known to be implicated in learning and plasticity include at least activity regulated cytoskeleton-associated protein (ArC), Homer1a and the AMPAr ($\alpha$-amino-3-hydroxyl-5-methyl-4-isoxazole-propionate receptor) subunit Glur1 \cite{Redondo2011}. This means that the variables of these phenomenological models should not be interpreted as concentrations of single molecules, but should be viewed as ``reporters" indicating important changes in the molecular configuration of the synapse (see the Discussion of \cite{Ziegler2015}). 

\subsection{Optimality}

The approximate $1/\sqrt{t}$ decay of the memory trace exhibited by the model in \cite{bf16} is the slowest allowed among power-law decays. Slower decays lead to synaptic efficacies that accumulate changes too rapidly and grow without bound. Interestingly, one can prove (see Suppl. Info. of \cite{bf16}) that the $1/\sqrt{t}$ decay maximizes the area between the log-log plot of the SNR and the threshold for memory retrieval (Fig. \ref{snr}).

This statement is true not only when one restricts the analysis to power laws, but also when all possible decay functions are considered. The rationale for maximizing the area under the log-log plot of the SNR can be summarized as follows: while we want to have a large SNR to be able to retrieve a memory from a weak cue (see \cite{Krauth1988,bf16} and the discussion above about the importance of large initial SNR), we do not want to spend all our resources making an already large SNR even larger. Thus we discount very large values by taking a logarithm. Similarly, while we want to achieve long memory lifetimes, we do not focus exclusively on this at the expense of severely diminishing the SNR, and therefore we also discount very long memory lifetimes by taking a logarithm. While putting less emphasis on extremely large signal to noise ratios and extremely long memory lifetimes is very plausible, the use of the logarithm as a discounting function is of course arbitrary. It is interesting to consider also the case in which the SNR is not discounted logarithmically, i.e. when one wants to maximize the area under the log-linear plot of the SNR. In this situation, the optimal decay is faster, namely $1/t$, as in some synaptic models \cite{rf13,fda05}.

\subsection{Best realistic models}

As discussed above, some of the synaptic models studied in the 80's exhibited a huge memory capacity because of the unrealistic assumption that the synaptic weights could vary in an unlimited range. For any reasonably realistic model all the dynamic variables should vary in a limited range and they cannot be modified with arbitrary precision. In \cite{lg13} the authors considered a very broad class of realistic models with binary synaptic weights and multiple discrete internal states. They used an elegant approach to derive an upper bound for the SNR that no realistic model can exceed. More specifically, they considered synaptic dynamics that can be described as a Markov chain. They assumed that the number of states $M$ of this Markov chain is finite, as required for any realistic model. The upper bound they derived starts at an initial SNR of order $\sqrt{N}$, where $N$ is the number of independent synapses, and from there slowly decays as an exponential $\sim \exp(-t/M)$ 
up to a number of memories of order $M$, after which it decays as a power law $\sim t^{-1}$. The upper bound was derived by determining the maximal SNR for every particular memory age. Hence it is not guaranteed that there exists a single model that has this upper bound as its SNR curve. The memory lifetime of a Markov chain model with $M$ internal states cannot exceed $\mathcal{O}(\sqrt{N} M)$ (i.e. order of $\sqrt{N}$).

These results indicate that one possible way to achieve a large SNR is to take advantage of biological complexity as in the bidirectional cascade model \cite{bf16}. Indeed when these models are discretized and described as Markov chains, the number of states $M$ can grow exponentially with the number of dynamical variables. Large $M$s can then be achieved even when each individual variable has a relatively small number of states (i.e. a realistically low precision). In the case of the bidirectional cascade model the number of variables and the number of states of each variable are required to grow with $N$, but very slowly (the number of variables should scale as $\log N$ and the number of states per variable scales at most as $\sqrt{\log N}$).

\subsection{The role of sparseness}

The estimates discussed in the previous sections are based on the assumption that the patterns of desirable synaptic modifications induced by stimulation are dense and most synapses are affected. This could be a reasonable assumption when relatively small neural circuits are considered, but in large networks it is likely that only a small fraction of the synapses are significantly modified to store a new memory. Sparse patterns of synaptic modifications can strongly reduce the interference between different memories, and hence lead to extended memory lifetimes. In the extreme case of completely orthogonal sparse representations, it is possible to construct neural network models for which there is no memory decay. However, also in this case the number of memories that can be stored has an upper bound that is determined by the number of neurons $N_n$, simply because for $N_n$ neurons the maximum number of orthogonal representations is $N_n$. A more interesting case, in which the number of storable memories can actually be larger, is the case of random uncorrelated memories whose neural representations are sparse,  i.e. with a
small fraction $f$ of active neurons \cite{w69,tf88,Treves1990,tr91,af94,bcf98,am02,bf07,lk06,lk08,Hawkins09,dab14,bf16}. For many reasonable learning rules, these neural representations imply that the pattern of synaptic modifications is also sparse (e.g. if the synapses connecting two active neurons are potentiated, then only a fraction $f^2$ of the synapses is modified). There are also situations in which sparseness can be achieved at the dendritic level \cite{wm09,Wu2019} and it does not require sparseness at the neural level.

In all these cases the memory lifetime can scale almost quadratically with the number $N_{n}$ of neurons when the representations are sparse enough (i.e.~when $f$, the average fraction of active neurons, scales approximately as $1/N_{n}$). This is a significant improvement over the linear scaling obtained for dense representations. However, this capacity increase entails a reduction in the amount of information stored per memory and in the initial SNR. Scaling properties of different models are summarized in Table \ref{t2}. Interestingly, all models previously discussed are strongly affected by sparseness.

The beneficial effects of sparseness that led to this improvement in memory performance are at least threefold: the first one is a reduction in the noise, which occurs under the assumption that during retrieval the pattern of activity imposed on the network reads out only the $f\,N_n$ synapses (selected by the $f\,N_n$ active neurons) that were potentially modified during the storage of the memory to be retrieved. The second one is the sparsification of the synaptic modifications, as for some learning rules it is possible to greatly reduce the number of synapses that are modified by the storage of each memory (the average fraction of modified synapses could be as low as $f^2$). This sparsification is almost equivalent to changing the learning rate, or to rescaling forgetting times by a factor of $1/f^2$. The third one is a reduction in the correlations between different synapses. This third benefit can be extremely important given that in many situations the synapses are correlated even when the neural patterns representing the memories are uncorrelated (e.g. the synapses on the same dendritic tree could be correlated simply because they share the same post-synaptic neuron\cite{af94,Savin2014}). These correlations can be highly disruptive and can compromise the favorable scaling properties discussed above.

It is important to remember that $f$ has to scale with the number of neurons of the circuit in order to achieve a superlinear scaling of the capacity. While $f \sim 1/N_n$ may be a reasonable assumption which is compatible with electrophysiological data when $N_n$ is the number of neurons of the local circuit, this is no longer true when we consider neural circuits of a significantly larger size. Moreover, sparseness can also be beneficial in terms of generalization (see e.g.\cite{olshausen2004sparse}), but only if $f$ is not too small \cite{brf13}. For these reasons, sparse representations are unlikely to be the sole solution to the memory problem. Nevertheless, plausible levels of sparsity can certainly increase the number of memories that can be stored, and this advantage can be combined with those of synaptic complexity.

\subsection{Sparseness and correlated memories}

Sparseness is typically assumed to be a property of the random uncorrelated neural representations that are considered for the estimates of memory capacity. However, it might also be the result of a pre-processing procedure that extracts a sparse uncorrelated component of memories which have a dense representation\cite{Gluck1993,schapiro2017,benna2019}. In our everyday experiences, most of the new memories are similar to previously stored ones. This is the typical situation in the case of semantic memories, which contain information about categorical and functional relationships between familiar objects. For this type of memories we can utilize our previous knowledge about the objects so that we can store only the information about the relations between them (see e.g.\cite{mcclelland95}). In other words, we can clearly take advantage of the correlations between the new memory and the previously stored ones that encode the relevant objects. An efficient way of storing these memories is to exploit all possible correlations of this type, and then store only the memory component whose information is incompressible. This component, containing less information than the whole memory, can be represented with a significantly sparser neural representation. Memories are probably actively and passively reorganized to separate the correlated and the sparse incompressible part of the storable information. Modeling this process of recoding or reorganization, which probably involves the hippocampus, is of fundamental importance and it has been subject of several theoretical studies which started with the pioneering work of David Marr in the 70's~\cite{marr71} and continued in the 80's and in the 90's with the first memory models of the hippocampus \cite{Mcnaughton1987,Gluck1993,OReilly1994,Treves1994,Mcclelland1996}. In these models the authors proposed neural network models in which the representations of memories are first orthogonalized in order to become more separable and hence facilitate the storage and reconstruction of memories. This orthogonalization process can be explicitly modeled as a process of compression~\cite{Gluck1993,oreillyfrank2006,Kali2004,Battaglia2011a,schapiro2017,benna2019,whittington2020}). This process is probably implemented implicitly also in complex networks like deep networks, when they are trained with algorithms like back-propagation \cite{Shwartz2017}.

\subsection{Temporal correlations and catastrophic forgetting}

So far we considered random and uncorrelated memories, and in the last paragraph, memories that are correlated to each other. In the real world the sequences of experiences we go through are also temporally correlated, often on multiple timescales. These temporal correlations are highly disruptive and lead to what is known as catastrophic forgetting \cite{McCloskey1989}. To illustrate the problem consider a neural network that is trained to classify handwritten digits, the MNIST dataset. If the digit samples of the dataset are presented in a random order, then even a simple feedforward network trained with backpropagation can reach a very high classification performance, close to 100\%. However, imagine now that the digits are presented in a particular order: for example in a first session the network sees only 0s and 1s. Then in a second session only 2s and 3s, and it never sees 0s and 1s again. After 5 sessions each with a pair of digits (this protocol is called split-MNIST), the network can classify accurately the last two digits seen, the 8s and 9s, but the performance for the other digits is close to chance. The digits of the early sessions have been completely forgotten, overwritten by those of the last session.

The sequences of the split-MNIST protocol exhibit very simple temporal correlations: e.g. the probability of presenting a 0 again is 0.5 within the first session and then it goes to 0 in the following sessions. This leads to an autocorrelation function that is different from zero only on timescales of the order of the length of the session. These very simply temporal correlations are sufficient to disrupt dramatically the performance of the network. This scenario in which memories are stored in a particular order that exhibits temporal correlations is what we often experience in everyday life. For example the statistics of the visual stimuli we see while driving a car are very different during the day and after dark, and they are correlated over timescales of a few hours. We never see day and night scene intermingled in a random order. Moreover, driving in winter in the snow can be very different from driving in the summer, and here the changes occur on an even longer timescale. Nevertheless when we learn to drive in the snow we do not forget how to drive on a clean street. This is a fundamental problem of backpropagation and it is known from the early days. The most efficient solutions are based on a simple idea: while we are learning we can store in memory some samples and replay them at a later time \cite{mcclelland95}. So in session 1 of the split-MNIST protocol we could save samples of 0s and 1s, and when we learn 2s and 3s we can interleave the saved samples at random times, so that the sequence of samples in session 2 is basically a sequence of 0s,1s,2s and 3s in a random order. Given that catastrophic forgetting is a fundamental problem in machine learning, the literature about it is vast and it would be impossible to cover it here. However, in the context of this chapter it is important to mention that there are recent biologically realistic implementations of the replay activity idea (see e.g.\cite{Van2020}) that highlight the importance of the hippocampus as a temporary memory storage system, used to store some samples in an explicit or implicit way. Other solutions assume complex synaptic models \cite{zenke2017, kirkpatrick2017} designed to protect the important synapses from overwriting.

\begin{table}[hp]
	\vspace{2.5cm}
	\begin{center}
		\begin{tabular}{| l | c | c | c | c | c | }
			\hline
									& {\bf Time dep. }       & {\bf Initial}   & {\bf Min.} & {\bf Memory } & {\bf Tot.} \\
									& {\bf of $\mathcal{S}$} & $\mathcal{S/N}$ & $f$        &{\bf lifetime} & {\bf info.}\\ \hline
			
			Unbounded               & const.           & n/a        & $1/N$        & $N^2$ & $N$ \\  \hline
			Bistable $\pm 1$ & $e^{-t/f^2}$     & $f\sqrt{N}$ & $1/\sqrt{N}$ & $N$   & $\sqrt{N}$ \\  \hline
			Bistable $0,1$   & $e^{-t/f^2}$     & $\sqrt{Nf}$ & $1/N$ & $N^2$ & $N$\\  \hline
			Cascade model $0,1$          & $1/( t f^2)$      & ${\sqrt{Nf}}$ & $1/N$ & $N^2$ & $N$ \\ \hline
			Bidirectional cascade model           & $1/\sqrt{t}$ & $\sqrt{{N} \over f}$ &$1/N$& $N^2$ & $N$   \\  \hline
		\end{tabular}
	\caption{\small Approximate scaling properties of different synaptic models in the case of sparse neural representations ($f$ is the average fraction of active neurons). $\mathcal{S}$ is the memory signal, the initial $\mathcal{S/N}$ is the memory strength immediately after a memory is stored, following the storage of an infinite number of random uncorrelated memories. Basically we consider the initial SNR of a typical memory, which is part of a long stream of stored memories. Memory lifetime is defined as the time at which the SNR goes below the memory retrieval threshold. The last column describes the total amount of information that is storable (the information per memory scales as $fN$). Min. $f$ indicates what is the smallest $f$ that allows for an initial SNR that is larger than 1. The memory lifetime and the total storable information are computed for the minimal $f$. Unbounded refers to model proposed in \protect\cite{tf88} in which the synaptic variables can vary in an unlimited range. As in the case of the Hopfield model, there is no steady state, so we do not report an initial SNR. Bistable synapses have two stable synaptic values and the transitions between them are stochastic \protect\cite{af94}. Synapses are fast for potentiation (the transition probability is order 1) and relatively slow for depression (the transition probability scales as $f$). The cascade model is described in \protect\cite{bf07} for the sparse case. The bidirectional cascade model in \protect\cite{bf16}.}
	\label{t2}
	\end{center}
\end{table}

\pagebreak

\section{Memory retrieval}
\label{memretrieval}

The information stored in the synaptic weights is read out every time a pattern of neural activity is imposed on the pre-synaptic neurons. The weighted activity of these neurons will then affect the dynamics of the neural circuit they belong to, eventually leading to a behavioral response or to a change in the internal state of the brain. All these processes can be considered as a form of memory retrieval. However, there are neural network models whose dynamics implement specific types of memory retrieval. Typically, memories are represented by patterns of neural activity that are stored in the synaptic weights. These patterns can be retrieved at a later time by stimulating the network with a memory cue that is similar to one of the stored memories. The prototypical model of this form of auto-associative memory is the Hopfield attractor neural network model \cite{hopfield82}, which has been highly influential especially among physicists who used statistical mechanics techniques to study its memory capacity (see e.g. \citeNP{ags85,a89}).

\subsection{Auto-associative memories: the Hopfield model}

In the Hopfield model the memories are random patterns of activity and the synapses are modified to ensure that each memory is a stable fixed point of the neural dynamics (an attractor). The retrieval process can be described as a process of relaxation to one of the stored memories. Typically the process starts by imposing on the network a memory cue, which is a pattern of activity that is similar to one of the stored memories. Each neuron can be either active or inactive, and it computes the thresholded weighted sum of the activities of the other neurons. If the number of memories is not too large, the dynamics relaxes to the closest attractor, which represents the retrieved memory. This process of relaxation into an attractor can be described using an energy function, which, in the absence of noise, decreases every time the state of activation of a neuron is updated. The synaptic matrix was designed to have a hilly energy landscape, in which the minima (the bottom of the valleys) correspond to the stored memories. The energy is related to the SNR that we introduced as a measure of the memory strength. The relation is really simply because the energy is  equal to -SNR. Hence, a strong memory (high SNR) corresponds to a deep valley, which is highly attractive. As a consequence, a high SNR enables the Hopfield network to actually retrieve the stored memory.

In this chapter we focus on memory capacity, however it is important to mention that the Hopfield model provides a nice framework for understanding many other aspects of associative memory. For example simple extensions of the original model can explain several observations in experiments on paired-associate learning \cite{Rizzuto2001,Mongillo2003}.

\subsection{Scaling properties of memory retrieval models}

In the original Hopfield model, which has unbounded synapses, random uncorrelated memories can be successfully retrieved if the number of stored memories is below some critical value, which can be computed using statistical mechanics techniques. The number of memories $p$ should be smaller than $0.14 N_n$, where $N_n$ is the number of neurons in the network. As discussed above, the memory capacity is limited by the noise, which is due to memory interference (the signal is constant, whereas the noise increases as $\sqrt{p}$). Notice that in the case of the Hopfield network the number of synapses is $N=N_n^2$ because the network is fully connected. Naively, the ideal observer approach that we discussed above would then predict that $p$ should scale as $N_n^2$.
However, it is important to remind that the ideal observer estimate is based on the assumption that the synapses are statistically independent, and this is not the case in the Hopfield network. Indeed, different neurons receive basically the same input (any two neurons share $N-2$ inputs). A set of statistically independent synapses would be those on the dendritic tree of any particular neuron (for independent inputs), and therefore no larger than the total number of neurons $N_n$. So, when independent synapses are counted ($N_n$ and not $N_n^2$) the ideal observer approach correctly predicts that $p$ scales linearly with $N_n$, the number of neurons.

There are versions of the Hopfield model with bounded and even binary synapses (see e.g. \cite{t90,af94}), attractor models with sparse representations (see e.g. \cite{tf88,a89,af94})  and even spiking neural networks \cite{am02} that exhibit the same scaling properties as those predicted by the ideal observer approach. The scaling properties of memory retrieval have also been tested for complex synapses\cite{bf16} both in feed-forward and recurrent networks. The complex synapses of \cite{bf16} have also been employed to solve a more realistic continual reinforcement learning problem \cite{kaplanis2018}. Finally, it is important to remember that most of the works listed until now focused on random and uncorrelated patterns, but there are several extensions of the Hopfield model that can deal with arbitrary, correlated and even non-linearly separable patterns (see e.g. \cite{ackley1985, rigotti10}).

It is important to stress that although all the optimal memory models share the same scaling properties, there are still significant differences in the actual number of memories that can be retrieved when different neural dynamics are considered. It is only in the recent years that investigators started to study systematically optimal dynamics for memory retrieval \cite{Amit2010,Savin2014}.

In all the works that we cited there is a nice correspondence between the memory strength estimated using the ideal observer approach and the ability to retrieve a memory.
However, there is an interesting case in which the memories are stored strongly but the number of memories that can be recalled is rather limited. We now discuss this case, which is free recall.

\subsection{Free recall: when strong memories cannot be recalled}

Free recall is a standard experimental paradigm used to test the ability to store and retrieve unstructured information (see e.g. \cite{Kahana2012}). In a typical experiment on free recall, a subject is briefly exposed to a list of randomly assembled words, and it is later asked to recall as many words as possible in an arbitrary order. When the presented list becomes longer, the average number of recalled words grows in a sublinear way \cite{Standing1973,Murray1976} and interestingly, is much smaller that the number of words that the subject can recognize as belonging to the list. Indeed, when the subjects are asked to report whether they have seen a particular word or not, the performance is significantly better, indicating that the information about whether a word was in the list or not is still in memory. 

Recently the group of Tsodyks proposed a simple explanation \cite{Romani2013} based on a model that can reproduce quantitatively a surprisingly large number of experimental observations. The main idea is that the words of the list to be recalled are represented in the brain by overlapping sparse neuronal ensembles. Tsodyks and colleagues assumed that these representations are random. When the subjects are asked to recall the next item, they would select the one with a largest overlap to the current one, excluding the item that was recalled on the previous step. After a certain number of transitions, typically significantly smaller than the total number of items in the list, this process begins to cycle over already visited items. After the cycle is reached, no new items can be recalled. In \cite{Romani2013} the authors showed that the average number of recalled items scales as a power-law function of the total number of items in the list, with an exponent that depends on sparseness parameter $f$. In the limit for very sparse representations ($f\ll 1$) the dependence on $f$ becomes very weak and basically the number of recalled items $R$ depends only on the total number of items $p$:
$$R=\sqrt{3\pi p/2}$$
In this case, the model does not have any free parameters that can be tuned to fit the experimental results, and nevertheless it could accurately predict the number of recalled item in experiments on free recall with long lists \cite{Naim2019}.

\section{Conclusions}

Memory is a complex phenomenon that in the biological brain involves numerous highly diverse biochemical mechanisms. Synaptic plasticity is certainly one of the important mechanisms that the brain employs to store memories and learn and it is not surprising that it has been extensively studied both in experiments and theoretical studies. In this Chapter we focused on the theoretical work on memory capacity and we showed how mathematical models can be used to estimate the number of memories that can be stored, preserved over time and then retrieved in a network with a certain number of neurons and synapses. By comparing the memory performance of different models it is possible to identify the model features and their biological counterparts that are important for achieving a large memory capacity. In particular, the studies we covered could help to find the features that guarantee that the huge number of synaptic resources that are available in the human brain are harnessed efficiently.

The theory showed that the complexity of the biological synapses is actually important when one considers that the synaptic weights cannot vary in an unlimited range and cannot be modified with arbitrary precision.
The models that we discussed predict that the synaptic dynamics should depend on a number of variables that can be as large as the number of biochemical processes that are directly or indirectly involved in memory consolidation. In particular, the theory shows that the history dependence of plasticity (metaplasticity), which is a natural consequence of the complex network of interactions between biochemical processes, is a component of synaptic dynamics that is fundamentally important for storing memories efficiently. This greatly complicates both the theoretical and the experimental studies on synaptic plasticity because the same long term change induction protocol might lead to completely different outcomes in different experiments. A low dimensional phenomenological model that describes faithfully a series of experiments might fail in describing important observations in a different situation. For this reason, we need a new approach to the study of synaptic plasticity, in which we try to consider induction protocols that imitate as much as possible the long and complex series of modifications that are caused by the storage of real world memories.

The theory also shows that specific features of the neural representations like sparseness, the correlations between memory representations, and the temporal correlations in sequences of memories can strongly affect memory capacity. Sparse representations can greatly increase the number of uncorrelated memories that can be stored, whether the synapses are realistic (bounded, and with limited precision) or not. However, the amount of information stored per memory decreases as the representations of these random memories become sparser. This is not an issue when memories are highly structured, and hence correlated to each other, as the amount of information per memory can be significantly lower than in the case of random uncorrelated memories. In all these cases it is highly beneficial to construct sparse compressed representations of the memories to be stored. This could be one of the roles of the hippocampus, which certainly plays a fundamental role in memory consolidation.

Finally, the theory shows that often it is possible to retrieve all the stored information, as in the case of the famous Hopfield model. However, there are also situations, as in free recall, in which a large number of memories can be stored, but only a few can be recalled. These elegant theoretical studies show that the number of recalled items can be accurately predicted by a model. This is particularly surprising given that the model does not have any parameter to be tuned to reproduce the experimental observations.

The theoretical memory models and the mathematical tools that we described in this Chapter have all been developed to solve a computational problem related to memory capacity. This approach allowed us to identify important computational principles that underlie the neural substrate of memory. For this reason, these models can be highly valuable for designing, analyzing and interpreting future experiments.

\section{Acknowledgements}

I am very grateful to M. Benna for many fundamental discussions, comments and corrections that greatly improved the quality of a previous version of the article. I also want to thank M. Kahana for numerous suggestions and comments. SF is supported by the Gatsby Charitable Foundation, the Simons Foundation, the Swartz foundation, the Kavli foundation and the NSF's NeuroNex program award DBI-1707398.

\bibliographystyle{apacite}
\bibliography{oxfordarticle}

\hyphenation{Post-Script Sprin-ger}
\begin{thebibliography}{}

\bibitem [\protect \citeauthoryear {%
Abraham%
\ \BBA {} Bear%
}{%
Abraham%
\ \BBA {} Bear%
}{%
{\protect \APACyear {1996}}%
}]{%
abraham1996}
\APACinsertmetastar {%
abraham1996}%
\begin{APACrefauthors}%
Abraham, W\BPBI C.%
\BCBT {}\ \BBA {} Bear, M\BPBI F.%
\end{APACrefauthors}%
\unskip\
\newblock
\APACrefYearMonthDay{1996}{}{}.
\newblock
{\BBOQ}\APACrefatitle {Metaplasticity: the plasticity of synaptic plasticity}
  {Metaplasticity: the plasticity of synaptic plasticity}.{\BBCQ}
\newblock
\APACjournalVolNumPages{Trends in neurosciences}{19}{4}{126--130}.
\PrintBackRefs{\CurrentBib}

\bibitem [\protect \citeauthoryear {%
Ackley%
, Hinton%
\BCBL {}\ \BBA {} Sejnowski%
}{%
Ackley%
\ \protect \BOthers {.}}{%
{\protect \APACyear {1985}}%
}]{%
ackley1985}
\APACinsertmetastar {%
ackley1985}%
\begin{APACrefauthors}%
Ackley, D\BPBI H.%
, Hinton, G\BPBI E.%
\BCBL {}\ \BBA {} Sejnowski, T\BPBI J.%
\end{APACrefauthors}%
\unskip\
\newblock
\APACrefYearMonthDay{1985}{}{}.
\newblock
{\BBOQ}\APACrefatitle {A learning algorithm for Boltzmann machines} {A learning
  algorithm for boltzmann machines}.{\BBCQ}
\newblock
\APACjournalVolNumPages{Cognitive science}{9}{1}{147--169}.
\PrintBackRefs{\CurrentBib}

\bibitem [\protect \citeauthoryear {%
D\BPBI J.~Amit%
}{%
D\BPBI J.~Amit%
}{%
{\protect \APACyear {1989}}%
}]{%
a89}
\APACinsertmetastar {%
a89}%
\begin{APACrefauthors}%
Amit, D\BPBI J.%
\end{APACrefauthors}%
\unskip\
\newblock
\APACrefYear{1989}.
\newblock
\APACrefbtitle {Modeling Brain Function} {Modeling brain function}.
\newblock
\APACaddressPublisher{}{Cambridge University Press, NY}.
\PrintBackRefs{\CurrentBib}

\bibitem [\protect \citeauthoryear {%
D\BPBI J.~Amit%
\ \BBA {} Fusi%
}{%
D\BPBI J.~Amit%
\ \BBA {} Fusi%
}{%
{\protect \APACyear {1992}}%
}]{%
af92}
\APACinsertmetastar {%
af92}%
\begin{APACrefauthors}%
Amit, D\BPBI J.%
\BCBT {}\ \BBA {} Fusi, S.%
\end{APACrefauthors}%
\unskip\
\newblock
\APACrefYearMonthDay{1992}{}{}.
\newblock
{\BBOQ}\APACrefatitle {Constraints on learning in dynamic synapses}
  {Constraints on learning in dynamic synapses}.{\BBCQ}
\newblock
\APACjournalVolNumPages{Network}{3}{}{443}.
\PrintBackRefs{\CurrentBib}

\bibitem [\protect \citeauthoryear {%
D\BPBI J.~Amit%
\ \BBA {} Fusi%
}{%
D\BPBI J.~Amit%
\ \BBA {} Fusi%
}{%
{\protect \APACyear {1994}}%
}]{%
af94}
\APACinsertmetastar {%
af94}%
\begin{APACrefauthors}%
Amit, D\BPBI J.%
\BCBT {}\ \BBA {} Fusi, S.%
\end{APACrefauthors}%
\unskip\
\newblock
\APACrefYearMonthDay{1994}{}{}.
\newblock
{\BBOQ}\APACrefatitle {Learning in neural networks with material synapses}
  {Learning in neural networks with material synapses}.{\BBCQ}
\newblock
\APACjournalVolNumPages{Neural Computation}{6}{5}{957--982}.
\PrintBackRefs{\CurrentBib}

\bibitem [\protect \citeauthoryear {%
D\BPBI J.~Amit%
, Gutfreund%
\BCBL {}\ \BBA {} Sompolinsky%
}{%
D\BPBI J.~Amit%
\ \protect \BOthers {.}}{%
{\protect \APACyear {1985}}%
}]{%
ags85}
\APACinsertmetastar {%
ags85}%
\begin{APACrefauthors}%
Amit, D\BPBI J.%
, Gutfreund, H.%
\BCBL {}\ \BBA {} Sompolinsky, H.%
\end{APACrefauthors}%
\unskip\
\newblock
\APACrefYearMonthDay{1985}{}{}.
\newblock
{\BBOQ}\APACrefatitle {Storing infinite numbers of patterns in a spin-glass
  model of neural networks} {Storing infinite numbers of patterns in a
  spin-glass model of neural networks}.{\BBCQ}
\newblock
\APACjournalVolNumPages{Phys.~Rev.~Lett.}{55}{}{1530-1531}.
\PrintBackRefs{\CurrentBib}

\bibitem [\protect \citeauthoryear {%
D\BPBI J.~Amit%
\ \BBA {} Mongillo%
}{%
D\BPBI J.~Amit%
\ \BBA {} Mongillo%
}{%
{\protect \APACyear {2003}}%
}]{%
am02}
\APACinsertmetastar {%
am02}%
\begin{APACrefauthors}%
Amit, D\BPBI J.%
\BCBT {}\ \BBA {} Mongillo, G.%
\end{APACrefauthors}%
\unskip\
\newblock
\APACrefYearMonthDay{2003}{}{}.
\newblock
{\BBOQ}\APACrefatitle {Spike-driven synaptic dynamics generating working memory
  states} {Spike-driven synaptic dynamics generating working memory
  states}.{\BBCQ}
\newblock
\APACjournalVolNumPages{Neural Computation}{15}{3}{565--596}.
\PrintBackRefs{\CurrentBib}

\bibitem [\protect \citeauthoryear {%
Y.~Amit%
\ \BBA {} Huang%
}{%
Y.~Amit%
\ \BBA {} Huang%
}{%
{\protect \APACyear {2010}}%
}]{%
Amit2010}
\APACinsertmetastar {%
Amit2010}%
\begin{APACrefauthors}%
Amit, Y.%
\BCBT {}\ \BBA {} Huang, Y.%
\end{APACrefauthors}%
\unskip\
\newblock
\APACrefYearMonthDay{2010}{}{}.
\newblock
{\BBOQ}\APACrefatitle {Precise capacity analysis in binary networks with
  multiple coding level inputs} {Precise capacity analysis in binary networks
  with multiple coding level inputs}.{\BBCQ}
\newblock
\APACjournalVolNumPages{Neural computation}{22}{3}{660--688}.
\PrintBackRefs{\CurrentBib}

\bibitem [\protect \citeauthoryear {%
Arbib%
\ \BBA {} Bonaiuto%
}{%
Arbib%
\ \BBA {} Bonaiuto%
}{%
{\protect \APACyear {2016}}%
}]{%
arbib16}
\APACinsertmetastar {%
arbib16}%
\begin{APACrefauthors}%
Arbib, M\BPBI A.%
\BCBT {}\ \BBA {} Bonaiuto, J\BPBI J.%
\end{APACrefauthors}%
\unskip\
\newblock
\APACrefYear{2016}.
\newblock
\APACrefbtitle {From Neuron to Cognition via Computational Neuroscience} {From
  neuron to cognition via computational neuroscience}.
\newblock
\APACaddressPublisher{}{MIT Press}.
\PrintBackRefs{\CurrentBib}

\bibitem [\protect \citeauthoryear {%
Barak%
, Rigotti%
\BCBL {}\ \BBA {} Fusi%
}{%
Barak%
\ \protect \BOthers {.}}{%
{\protect \APACyear {2013}}%
}]{%
brf13}
\APACinsertmetastar {%
brf13}%
\begin{APACrefauthors}%
Barak, O.%
, Rigotti, M.%
\BCBL {}\ \BBA {} Fusi, S.%
\end{APACrefauthors}%
\unskip\
\newblock
\APACrefYearMonthDay{2013}{}{}.
\newblock
{\BBOQ}\APACrefatitle {The sparseness of mixed selectivity neurons controls the
  generalization--discrimination trade-off} {The sparseness of mixed
  selectivity neurons controls the generalization--discrimination
  trade-off}.{\BBCQ}
\newblock
\APACjournalVolNumPages{The Journal of Neuroscience}{33}{9}{3844--3856}.
\PrintBackRefs{\CurrentBib}

\bibitem [\protect \citeauthoryear {%
Barrett%
, Billings%
, Morris%
\BCBL {}\ \BBA {} Van~Rossum%
}{%
Barrett%
\ \protect \BOthers {.}}{%
{\protect \APACyear {2009}}%
}]{%
Barrett2009}
\APACinsertmetastar {%
Barrett2009}%
\begin{APACrefauthors}%
Barrett, A\BPBI B.%
, Billings, G\BPBI O.%
, Morris, R\BPBI G.%
\BCBL {}\ \BBA {} Van~Rossum, M\BPBI C.%
\end{APACrefauthors}%
\unskip\
\newblock
\APACrefYearMonthDay{2009}{}{}.
\newblock
{\BBOQ}\APACrefatitle {State based model of long-term potentiation and synaptic
  tagging and capture} {State based model of long-term potentiation and
  synaptic tagging and capture}.{\BBCQ}
\newblock
\APACjournalVolNumPages{PLoS Comput Biol}{5}{1}{e1000259}.
\PrintBackRefs{\CurrentBib}

\bibitem [\protect \citeauthoryear {%
Bartol~Jr%
\ \protect \BOthers {.}}{%
Bartol~Jr%
\ \protect \BOthers {.}}{%
{\protect \APACyear {2015}}%
}]{%
BartolJr2015}
\APACinsertmetastar {%
BartolJr2015}%
\begin{APACrefauthors}%
Bartol~Jr, T\BPBI M.%
, Bromer, C.%
, Kinney, J.%
, Chirillo, M\BPBI A.%
, Bourne, J\BPBI N.%
, Harris, K\BPBI M.%
\BCBL {}\ \BBA {} Sejnowski, T\BPBI J.%
\end{APACrefauthors}%
\unskip\
\newblock
\APACrefYearMonthDay{2015}{}{}.
\newblock
{\BBOQ}\APACrefatitle {Nanoconnectomic upper bound on the variability of
  synaptic plasticity} {Nanoconnectomic upper bound on the variability of
  synaptic plasticity}.{\BBCQ}
\newblock
\APACjournalVolNumPages{Elife}{4}{}{e10778}.
\PrintBackRefs{\CurrentBib}

\bibitem [\protect \citeauthoryear {%
Battaglia%
\ \BBA {} Pennartz%
}{%
Battaglia%
\ \BBA {} Pennartz%
}{%
{\protect \APACyear {2011}}%
}]{%
Battaglia2011a}
\APACinsertmetastar {%
Battaglia2011a}%
\begin{APACrefauthors}%
Battaglia, F\BPBI P.%
\BCBT {}\ \BBA {} Pennartz, C\BPBI M\BPBI A.%
\end{APACrefauthors}%
\unskip\
\newblock
\APACrefYearMonthDay{2011}{}{}.
\newblock
{\BBOQ}\APACrefatitle {The construction of semantic memory: grammar-based
  representations learned from relational episodic information.} {The
  construction of semantic memory: grammar-based representations learned from
  relational episodic information.}{\BBCQ}
\newblock
\APACjournalVolNumPages{Front Comput Neurosci}{5}{}{36}.
\newblock
\begin{APACrefURL} \url{http://dx.doi.org/10.3389/fncom.2011.00036}
  \end{APACrefURL}
\newblock
\begin{APACrefDOI} \doi{10.3389/fncom.2011.00036} \end{APACrefDOI}
\PrintBackRefs{\CurrentBib}

\bibitem [\protect \citeauthoryear {%
Ben Dayan~Rubin%
\ \BBA {} Fusi%
}{%
Ben Dayan~Rubin%
\ \BBA {} Fusi%
}{%
{\protect \APACyear {2007}}%
}]{%
bf07}
\APACinsertmetastar {%
bf07}%
\begin{APACrefauthors}%
Ben Dayan~Rubin, D\BPBI D.%
\BCBT {}\ \BBA {} Fusi, S.%
\end{APACrefauthors}%
\unskip\
\newblock
\APACrefYearMonthDay{2007}{}{}.
\newblock
{\BBOQ}\APACrefatitle {{{L}ong memory lifetimes require complex synapses and
  limited sparseness}} {{{L}ong memory lifetimes require complex synapses and
  limited sparseness}}.{\BBCQ}
\newblock
\APACjournalVolNumPages{Front Comput Neurosci}{1}{}{7}.
\PrintBackRefs{\CurrentBib}

\bibitem [\protect \citeauthoryear {%
Benna%
\ \BBA {} Fusi%
}{%
Benna%
\ \BBA {} Fusi%
}{%
{\protect \APACyear {2016}}%
}]{%
bf16}
\APACinsertmetastar {%
bf16}%
\begin{APACrefauthors}%
Benna, M\BPBI K.%
\BCBT {}\ \BBA {} Fusi, S.%
\end{APACrefauthors}%
\unskip\
\newblock
\APACrefYearMonthDay{2016}{}{}.
\newblock
{\BBOQ}\APACrefatitle {Computational principles of synaptic memory
  consolidation} {Computational principles of synaptic memory
  consolidation}.{\BBCQ}
\newblock
\APACjournalVolNumPages{Nature neuroscience}{}{}{}.
\PrintBackRefs{\CurrentBib}

\bibitem [\protect \citeauthoryear {%
Benna%
\ \BBA {} Fusi%
}{%
Benna%
\ \BBA {} Fusi%
}{%
{\protect \APACyear {2021}}%
}]{%
benna2019}
\APACinsertmetastar {%
benna2019}%
\begin{APACrefauthors}%
Benna, M\BPBI K.%
\BCBT {}\ \BBA {} Fusi, S.%
\end{APACrefauthors}%
\unskip\
\newblock
\APACrefYearMonthDay{2021}{}{}.
\newblock
{\BBOQ}\APACrefatitle {Place cells may simply be memory cells: Memory
  compression leads to spatial tuning and history dependence.} {Place cells may
  simply be memory cells: Memory compression leads to spatial tuning and
  history dependence.}{\BBCQ}
\newblock
\APACjournalVolNumPages{Proceedings of the National Academy of Sciences of the
  United States of America}{118 51}{}{}.
\PrintBackRefs{\CurrentBib}

\bibitem [\protect \citeauthoryear {%
Bernardi%
\ \protect \BOthers {.}}{%
Bernardi%
\ \protect \BOthers {.}}{%
{\protect \APACyear {2020}}%
}]{%
Bernardi2020}
\APACinsertmetastar {%
Bernardi2020}%
\begin{APACrefauthors}%
Bernardi, S.%
, Benna, M\BPBI K.%
, Rigotti, M.%
, Munuera, J.%
, Fusi, S.%
\BCBL {}\ \BBA {} Salzman, C\BPBI D.%
\end{APACrefauthors}%
\unskip\
\newblock
\APACrefYearMonthDay{2020}{}{}.
\newblock
{\BBOQ}\APACrefatitle {The geometry of abstraction in the hippocampus and
  prefrontal cortex} {The geometry of abstraction in the hippocampus and
  prefrontal cortex}.{\BBCQ}
\newblock
\APACjournalVolNumPages{Cell}{183}{4}{954--967}.
\PrintBackRefs{\CurrentBib}

\bibitem [\protect \citeauthoryear {%
Block%
}{%
Block%
}{%
{\protect \APACyear {1962}}%
}]{%
block1962}
\APACinsertmetastar {%
block1962}%
\begin{APACrefauthors}%
Block, H\BHBI D.%
\end{APACrefauthors}%
\unskip\
\newblock
\APACrefYearMonthDay{1962}{}{}.
\newblock
{\BBOQ}\APACrefatitle {The perceptron: A model for brain functioning. i} {The
  perceptron: A model for brain functioning. i}.{\BBCQ}
\newblock
\APACjournalVolNumPages{Reviews of Modern Physics}{34}{1}{123}.
\PrintBackRefs{\CurrentBib}

\bibitem [\protect \citeauthoryear {%
Brunel%
, Carusi%
\BCBL {}\ \BBA {} Fusi%
}{%
Brunel%
\ \protect \BOthers {.}}{%
{\protect \APACyear {1998}}%
}]{%
bcf98}
\APACinsertmetastar {%
bcf98}%
\begin{APACrefauthors}%
Brunel, N.%
, Carusi, F.%
\BCBL {}\ \BBA {} Fusi, S.%
\end{APACrefauthors}%
\unskip\
\newblock
\APACrefYearMonthDay{1998}{Feb}{}.
\newblock
{\BBOQ}\APACrefatitle {{Slow stochastic Hebbian learning of classes of stimuli
  in a recurrent neural network}} {{Slow stochastic Hebbian learning of classes
  of stimuli in a recurrent neural network}}.{\BBCQ}
\newblock
\APACjournalVolNumPages{Network}{9}{1}{123--152}.
\PrintBackRefs{\CurrentBib}

\bibitem [\protect \citeauthoryear {%
Carpenter%
\ \BBA {} Grossberg%
}{%
Carpenter%
\ \BBA {} Grossberg%
}{%
{\protect \APACyear {1991}}%
}]{%
cg91}
\APACinsertmetastar {%
cg91}%
\begin{APACrefauthors}%
Carpenter, G.%
\BCBT {}\ \BBA {} Grossberg, S.%
\end{APACrefauthors}%
\unskip\
\newblock
\APACrefYear{1991}.
\newblock
\APACrefbtitle {Pattern Recognition by Self-Organizing Neural Networks}
  {Pattern recognition by self-organizing neural networks}.
\newblock
\APACaddressPublisher{}{MIT Press}.
\PrintBackRefs{\CurrentBib}

\bibitem [\protect \citeauthoryear {%
Chaudhuri%
\ \BBA {} Fiete%
}{%
Chaudhuri%
\ \BBA {} Fiete%
}{%
{\protect \APACyear {2016}}%
}]{%
fiete16}
\APACinsertmetastar {%
fiete16}%
\begin{APACrefauthors}%
Chaudhuri, R.%
\BCBT {}\ \BBA {} Fiete, I.%
\end{APACrefauthors}%
\unskip\
\newblock
\APACrefYearMonthDay{2016}{}{}.
\newblock
{\BBOQ}\APACrefatitle {Computational principles of memory.} {Computational
  principles of memory.}{\BBCQ}
\newblock
\APACjournalVolNumPages{Nature Neuroscience}{}{}{}.
\PrintBackRefs{\CurrentBib}

\bibitem [\protect \citeauthoryear {%
Clopath%
, Ziegler%
, Vasilaki%
, Busing%
\BCBL {}\ \BBA {} Gerstner%
}{%
Clopath%
\ \protect \BOthers {.}}{%
{\protect \APACyear {2008}}%
}]{%
clopath08}
\APACinsertmetastar {%
clopath08}%
\begin{APACrefauthors}%
Clopath, C.%
, Ziegler, L.%
, Vasilaki, E.%
, Busing, L.%
\BCBL {}\ \BBA {} Gerstner, W.%
\end{APACrefauthors}%
\unskip\
\newblock
\APACrefYearMonthDay{2008}{Dec}{}.
\newblock
{\BBOQ}\APACrefatitle {{{T}ag-trigger-consolidation: a model of early and late
  long-term-potentiation and depression}} {{{T}ag-trigger-consolidation: a
  model of early and late long-term-potentiation and depression}}.{\BBCQ}
\newblock
\APACjournalVolNumPages{PLoS Comput. Biol.}{4}{}{e1000248}.
\PrintBackRefs{\CurrentBib}

\bibitem [\protect \citeauthoryear {%
Crick%
}{%
Crick%
}{%
{\protect \APACyear {1984}}%
}]{%
crick1984}
\APACinsertmetastar {%
crick1984}%
\begin{APACrefauthors}%
Crick, F.%
\end{APACrefauthors}%
\unskip\
\newblock
\APACrefYearMonthDay{1984}{}{}.
\newblock
{\BBOQ}\APACrefatitle {{{M}emory and molecular turnover}} {{{M}emory and
  molecular turnover}}.{\BBCQ}
\newblock
\APACjournalVolNumPages{Nature}{312}{}{101}.
\PrintBackRefs{\CurrentBib}

\bibitem [\protect \citeauthoryear {%
Dubreuil%
, Amit%
\BCBL {}\ \BBA {} Brunel%
}{%
Dubreuil%
\ \protect \BOthers {.}}{%
{\protect \APACyear {2014}}%
}]{%
dab14}
\APACinsertmetastar {%
dab14}%
\begin{APACrefauthors}%
Dubreuil, A\BPBI M.%
, Amit, Y.%
\BCBL {}\ \BBA {} Brunel, N.%
\end{APACrefauthors}%
\unskip\
\newblock
\APACrefYearMonthDay{2014}{}{}.
\newblock
{\BBOQ}\APACrefatitle {Memory capacity of networks with stochastic binary
  synapses} {Memory capacity of networks with stochastic binary
  synapses}.{\BBCQ}
\newblock
\APACjournalVolNumPages{PLoS Comput Biol}{10}{8}{e1003727}.
\PrintBackRefs{\CurrentBib}

\bibitem [\protect \citeauthoryear {%
Emes%
\ \BBA {} Grant%
}{%
Emes%
\ \BBA {} Grant%
}{%
{\protect \APACyear {2012}}%
}]{%
EmesGrant2012}
\APACinsertmetastar {%
EmesGrant2012}%
\begin{APACrefauthors}%
Emes, R\BPBI D.%
\BCBT {}\ \BBA {} Grant, S\BPBI G.%
\end{APACrefauthors}%
\unskip\
\newblock
\APACrefYearMonthDay{2012}{}{}.
\newblock
{\BBOQ}\APACrefatitle {Evolution of synapse complexity and diversity}
  {Evolution of synapse complexity and diversity}.{\BBCQ}
\newblock
\APACjournalVolNumPages{Annual review of neuroscience}{35}{}{111--131}.
\PrintBackRefs{\CurrentBib}

\bibitem [\protect \citeauthoryear {%
Fusi%
}{%
Fusi%
}{%
{\protect \APACyear {2002}}%
}]{%
f02}
\APACinsertmetastar {%
f02}%
\begin{APACrefauthors}%
Fusi, S.%
\end{APACrefauthors}%
\unskip\
\newblock
\APACrefYearMonthDay{2002}{}{}.
\newblock
{\BBOQ}\APACrefatitle {{Hebbian spike-driven synaptic plasticity for learning
  patterns of mean firing rates}} {{Hebbian spike-driven synaptic plasticity
  for learning patterns of mean firing rates}}.{\BBCQ}
\newblock
\APACjournalVolNumPages{Biol Cybern}{87}{5-6}{459--470}.
\PrintBackRefs{\CurrentBib}

\bibitem [\protect \citeauthoryear {%
Fusi%
\ \BBA {} Abbott%
}{%
Fusi%
\ \BBA {} Abbott%
}{%
{\protect \APACyear {2007}}%
}]{%
fa07}
\APACinsertmetastar {%
fa07}%
\begin{APACrefauthors}%
Fusi, S.%
\BCBT {}\ \BBA {} Abbott, L\BPBI F.%
\end{APACrefauthors}%
\unskip\
\newblock
\APACrefYearMonthDay{2007}{}{}.
\newblock
{\BBOQ}\APACrefatitle {{{L}imits on the memory storage capacity of bounded
  synapses}} {{{L}imits on the memory storage capacity of bounded
  synapses}}.{\BBCQ}
\newblock
\APACjournalVolNumPages{Nat. Neurosci.}{10}{}{485--493}.
\PrintBackRefs{\CurrentBib}

\bibitem [\protect \citeauthoryear {%
Fusi%
, Asaad%
, Miller%
\BCBL {}\ \BBA {} Wang%
}{%
Fusi%
\ \protect \BOthers {.}}{%
{\protect \APACyear {2007}}%
}]{%
Fusi2007}
\APACinsertmetastar {%
Fusi2007}%
\begin{APACrefauthors}%
Fusi, S.%
, Asaad, W\BPBI F.%
, Miller, E\BPBI K.%
\BCBL {}\ \BBA {} Wang, X\BHBI J.%
\end{APACrefauthors}%
\unskip\
\newblock
\APACrefYearMonthDay{2007}{}{}.
\newblock
{\BBOQ}\APACrefatitle {A neural circuit model of flexible sensorimotor mapping:
  learning and forgetting on multiple timescales} {A neural circuit model of
  flexible sensorimotor mapping: learning and forgetting on multiple
  timescales}.{\BBCQ}
\newblock
\APACjournalVolNumPages{Neuron}{54}{2}{319--333}.
\PrintBackRefs{\CurrentBib}

\bibitem [\protect \citeauthoryear {%
Fusi%
, Drew%
\BCBL {}\ \BBA {} Abbott%
}{%
Fusi%
\ \protect \BOthers {.}}{%
{\protect \APACyear {2005}}%
}]{%
fda05}
\APACinsertmetastar {%
fda05}%
\begin{APACrefauthors}%
Fusi, S.%
, Drew, P.%
\BCBL {}\ \BBA {} Abbott, L\BPBI F.%
\end{APACrefauthors}%
\unskip\
\newblock
\APACrefYearMonthDay{2005}{}{}.
\newblock
{\BBOQ}\APACrefatitle {{Cascade models of synaptically stored memories}}
  {{Cascade models of synaptically stored memories}}.{\BBCQ}
\newblock
\APACjournalVolNumPages{Neuron}{45}{4}{599--611}.
\PrintBackRefs{\CurrentBib}

\bibitem [\protect \citeauthoryear {%
George%
\ \BBA {} Hawkins%
}{%
George%
\ \BBA {} Hawkins%
}{%
{\protect \APACyear {2009}}%
}]{%
Hawkins09}
\APACinsertmetastar {%
Hawkins09}%
\begin{APACrefauthors}%
George, D.%
\BCBT {}\ \BBA {} Hawkins, J.%
\end{APACrefauthors}%
\unskip\
\newblock
\APACrefYearMonthDay{2009}{}{}.
\newblock
{\BBOQ}\APACrefatitle {Towards a mathematical theory of cortical
  micro-circuits} {Towards a mathematical theory of cortical
  micro-circuits}.{\BBCQ}
\newblock
\APACjournalVolNumPages{PLoS Comput Biol}{5}{10}{e1000532}.
\PrintBackRefs{\CurrentBib}

\bibitem [\protect \citeauthoryear {%
Gluck%
\ \BBA {} Myers%
}{%
Gluck%
\ \BBA {} Myers%
}{%
{\protect \APACyear {1993}}%
}]{%
Gluck1993}
\APACinsertmetastar {%
Gluck1993}%
\begin{APACrefauthors}%
Gluck, M\BPBI A.%
\BCBT {}\ \BBA {} Myers, C\BPBI E.%
\end{APACrefauthors}%
\unskip\
\newblock
\APACrefYearMonthDay{1993}{}{}.
\newblock
{\BBOQ}\APACrefatitle {Hippocampal mediation of stimulus representation: A
  computational theory} {Hippocampal mediation of stimulus representation: A
  computational theory}.{\BBCQ}
\newblock
\APACjournalVolNumPages{Hippocampus}{3}{4}{491--516}.
\PrintBackRefs{\CurrentBib}

\bibitem [\protect \citeauthoryear {%
Hebb%
}{%
Hebb%
}{%
{\protect \APACyear {1949}}%
}]{%
hebb49}
\APACinsertmetastar {%
hebb49}%
\begin{APACrefauthors}%
Hebb, D\BPBI O.%
\end{APACrefauthors}%
\unskip\
\newblock
\APACrefYear{1949}.
\newblock
\APACrefbtitle {Organization of behavior} {Organization of behavior}.
\newblock
\APACaddressPublisher{}{New York: Wiley}.
\PrintBackRefs{\CurrentBib}

\bibitem [\protect \citeauthoryear {%
Hertz%
, Krogh%
\BCBL {}\ \BBA {} Palmer%
}{%
Hertz%
\ \protect \BOthers {.}}{%
{\protect \APACyear {1991}}%
}]{%
hkp91}
\APACinsertmetastar {%
hkp91}%
\begin{APACrefauthors}%
Hertz, J.%
, Krogh, A.%
\BCBL {}\ \BBA {} Palmer, R.%
\end{APACrefauthors}%
\unskip\
\newblock
\APACrefYear{1991}.
\newblock
\APACrefbtitle {Introduction to the Theory of neural computation} {Introduction
  to the theory of neural computation}.
\newblock
\APACaddressPublisher{}{Addison Wesley}.
\PrintBackRefs{\CurrentBib}

\bibitem [\protect \citeauthoryear {%
Hopfield%
}{%
Hopfield%
}{%
{\protect \APACyear {1982}}%
{\protect \APACexlab {{\protect \BCnt {1}}}}}]{%
hopfield82}
\APACinsertmetastar {%
hopfield82}%
\begin{APACrefauthors}%
Hopfield, J\BPBI J.%
\end{APACrefauthors}%
\unskip\
\newblock
\APACrefYearMonthDay{1982{\protect \BCnt {1}}}{}{}.
\newblock
{\BBOQ}\APACrefatitle {Neural networks and physical systems with emergent
  collective computational abilities} {Neural networks and physical systems
  with emergent collective computational abilities}.{\BBCQ}
\newblock
\APACjournalVolNumPages{Proc.~Natl.~Acad.~Sci.~(USA)}{79}{}{2554-2558}.
\PrintBackRefs{\CurrentBib}

\bibitem [\protect \citeauthoryear {%
Hopfield%
}{%
Hopfield%
}{%
{\protect \APACyear {1982}}%
{\protect \APACexlab {{\protect \BCnt {2}}}}}]{%
h82}
\APACinsertmetastar {%
h82}%
\begin{APACrefauthors}%
Hopfield, J\BPBI J.%
\end{APACrefauthors}%
\unskip\
\newblock
\APACrefYearMonthDay{1982{\protect \BCnt {2}}}{}{}.
\newblock
{\BBOQ}\APACrefatitle {Neural networks and physical systems with emergent
  selective computational abilities} {Neural networks and physical systems with
  emergent selective computational abilities}.{\BBCQ}
\newblock
\APACjournalVolNumPages{Proc. Natl. Acad. Sci. USA}{79}{}{2554}.
\PrintBackRefs{\CurrentBib}

\bibitem [\protect \citeauthoryear {%
Iigaya%
\ \protect \BOthers {.}}{%
Iigaya%
\ \protect \BOthers {.}}{%
{\protect \APACyear {2019}}%
}]{%
iigaya2019}
\APACinsertmetastar {%
iigaya2019}%
\begin{APACrefauthors}%
Iigaya, K.%
, Ahmadian, Y.%
, Sugrue, L\BPBI P.%
, Corrado, G\BPBI S.%
, Loewenstein, Y.%
, Newsome, W\BPBI T.%
\BCBL {}\ \BBA {} Fusi, S.%
\end{APACrefauthors}%
\unskip\
\newblock
\APACrefYearMonthDay{2019}{}{}.
\newblock
{\BBOQ}\APACrefatitle {Deviation from the matching law reflects an optimal
  strategy involving learning over multiple timescales} {Deviation from the
  matching law reflects an optimal strategy involving learning over multiple
  timescales}.{\BBCQ}
\newblock
\APACjournalVolNumPages{Nature communications}{10}{1}{1--14}.
\PrintBackRefs{\CurrentBib}

\bibitem [\protect \citeauthoryear {%
Kahana%
}{%
Kahana%
}{%
{\protect \APACyear {2012}}%
}]{%
Kahana2012}
\APACinsertmetastar {%
Kahana2012}%
\begin{APACrefauthors}%
Kahana, M\BPBI J.%
\end{APACrefauthors}%
\unskip\
\newblock
\APACrefYear{2012}.
\newblock
\APACrefbtitle {Foundations of human memory} {Foundations of human memory}.
\newblock
\APACaddressPublisher{}{OUP USA}.
\PrintBackRefs{\CurrentBib}

\bibitem [\protect \citeauthoryear {%
Kali%
\ \BBA {} Dayan%
}{%
Kali%
\ \BBA {} Dayan%
}{%
{\protect \APACyear {2004}}%
}]{%
Kali2004}
\APACinsertmetastar {%
Kali2004}%
\begin{APACrefauthors}%
Kali, S.%
\BCBT {}\ \BBA {} Dayan, P.%
\end{APACrefauthors}%
\unskip\
\newblock
\APACrefYearMonthDay{2004}{Mar}{}.
\newblock
{\BBOQ}\APACrefatitle {Off-line replay maintains declarative memories in a
  model of hippocampal-neocortical interactions.} {Off-line replay maintains
  declarative memories in a model of hippocampal-neocortical
  interactions.}{\BBCQ}
\newblock
\APACjournalVolNumPages{Nat Neurosci}{7}{3}{286--294}.
\newblock
\begin{APACrefURL} \url{http://dx.doi.org/10.1038/nn1202} \end{APACrefURL}
\newblock
\begin{APACrefDOI} \doi{10.1038/nn1202} \end{APACrefDOI}
\PrintBackRefs{\CurrentBib}

\bibitem [\protect \citeauthoryear {%
Kaplanis%
, Shanahan%
\BCBL {}\ \BBA {} Clopath%
}{%
Kaplanis%
\ \protect \BOthers {.}}{%
{\protect \APACyear {2018}}%
}]{%
kaplanis2018}
\APACinsertmetastar {%
kaplanis2018}%
\begin{APACrefauthors}%
Kaplanis, C.%
, Shanahan, M.%
\BCBL {}\ \BBA {} Clopath, C.%
\end{APACrefauthors}%
\unskip\
\newblock
\APACrefYearMonthDay{2018}{}{}.
\newblock
{\BBOQ}\APACrefatitle {Continual reinforcement learning with complex synapses}
  {Continual reinforcement learning with complex synapses}.{\BBCQ}
\newblock
\APACjournalVolNumPages{arXiv preprint arXiv:1802.07239}{}{}{}.
\PrintBackRefs{\CurrentBib}

\bibitem [\protect \citeauthoryear {%
Kirkpatrick%
\ \protect \BOthers {.}}{%
Kirkpatrick%
\ \protect \BOthers {.}}{%
{\protect \APACyear {2017}}%
}]{%
kirkpatrick2017}
\APACinsertmetastar {%
kirkpatrick2017}%
\begin{APACrefauthors}%
Kirkpatrick, J.%
, Pascanu, R.%
, Rabinowitz, N.%
, Veness, J.%
, Desjardins, G.%
, Rusu, A\BPBI A.%
\BDBL {}others%
\end{APACrefauthors}%
\unskip\
\newblock
\APACrefYearMonthDay{2017}{}{}.
\newblock
{\BBOQ}\APACrefatitle {Overcoming catastrophic forgetting in neural networks}
  {Overcoming catastrophic forgetting in neural networks}.{\BBCQ}
\newblock
\APACjournalVolNumPages{Proceedings of the national academy of
  sciences}{114}{13}{3521--3526}.
\PrintBackRefs{\CurrentBib}

\bibitem [\protect \citeauthoryear {%
Krauth%
, M{\'e}zard%
\BCBL {}\ \BBA {} Nadal%
}{%
Krauth%
\ \protect \BOthers {.}}{%
{\protect \APACyear {1988}}%
}]{%
Krauth1988}
\APACinsertmetastar {%
Krauth1988}%
\begin{APACrefauthors}%
Krauth, W.%
, M{\'e}zard, M.%
\BCBL {}\ \BBA {} Nadal, J\BHBI P.%
\end{APACrefauthors}%
\unskip\
\newblock
\APACrefYearMonthDay{1988}{}{}.
\newblock
{\BBOQ}\APACrefatitle {Basins of attraction in a perceptron-like neural
  network} {Basins of attraction in a perceptron-like neural network}.{\BBCQ}
\newblock
\APACjournalVolNumPages{Complex Systems}{2}{4}{387--408}.
\PrintBackRefs{\CurrentBib}

\bibitem [\protect \citeauthoryear {%
Kumaran%
, Hassabis%
\BCBL {}\ \BBA {} McClelland%
}{%
Kumaran%
\ \protect \BOthers {.}}{%
{\protect \APACyear {2016}}%
}]{%
KUMARAN2016}
\APACinsertmetastar {%
KUMARAN2016}%
\begin{APACrefauthors}%
Kumaran, D.%
, Hassabis, D.%
\BCBL {}\ \BBA {} McClelland, J\BPBI L.%
\end{APACrefauthors}%
\unskip\
\newblock
\APACrefYearMonthDay{2016}{}{}.
\newblock
{\BBOQ}\APACrefatitle {What Learning Systems do Intelligent Agents Need?
  Complementary Learning Systems Theory Updated} {What learning systems do
  intelligent agents need? complementary learning systems theory
  updated}.{\BBCQ}
\newblock
\APACjournalVolNumPages{Trends in Cognitive Sciences}{20}{7}{512-534}.
\PrintBackRefs{\CurrentBib}

\bibitem [\protect \citeauthoryear {%
Lahiri%
\ \BBA {} Ganguli%
}{%
Lahiri%
\ \BBA {} Ganguli%
}{%
{\protect \APACyear {2013}}%
}]{%
lg13}
\APACinsertmetastar {%
lg13}%
\begin{APACrefauthors}%
Lahiri, S.%
\BCBT {}\ \BBA {} Ganguli, S.%
\end{APACrefauthors}%
\unskip\
\newblock
\APACrefYearMonthDay{2013}{}{}.
\newblock
{\BBOQ}\APACrefatitle {A memory frontier for complex synapses} {A memory
  frontier for complex synapses}.{\BBCQ}
\newblock
\BIn{} \APACrefbtitle {Advances in Neural Information Processing Systems}
  {Advances in neural information processing systems}\ (\BPGS\ 1034--1042).
\PrintBackRefs{\CurrentBib}

\bibitem [\protect \citeauthoryear {%
LeCun%
, Bengio%
\BCBL {}\ \BBA {} Hinton%
}{%
LeCun%
\ \protect \BOthers {.}}{%
{\protect \APACyear {2015}}%
}]{%
LeCun2015}
\APACinsertmetastar {%
LeCun2015}%
\begin{APACrefauthors}%
LeCun, Y.%
, Bengio, Y.%
\BCBL {}\ \BBA {} Hinton, G.%
\end{APACrefauthors}%
\unskip\
\newblock
\APACrefYearMonthDay{2015}{}{}.
\newblock
{\BBOQ}\APACrefatitle {Deep learning} {Deep learning}.{\BBCQ}
\newblock
\APACjournalVolNumPages{Nature}{521}{7553}{436--444}.
\PrintBackRefs{\CurrentBib}

\bibitem [\protect \citeauthoryear {%
Leibold%
\ \BBA {} Kempter%
}{%
Leibold%
\ \BBA {} Kempter%
}{%
{\protect \APACyear {2006}}%
}]{%
lk06}
\APACinsertmetastar {%
lk06}%
\begin{APACrefauthors}%
Leibold, C.%
\BCBT {}\ \BBA {} Kempter, R.%
\end{APACrefauthors}%
\unskip\
\newblock
\APACrefYearMonthDay{2006}{}{}.
\newblock
{\BBOQ}\APACrefatitle {Memory capacity for sequences in a recurrent network
  with biological constraints} {Memory capacity for sequences in a recurrent
  network with biological constraints}.{\BBCQ}
\newblock
\APACjournalVolNumPages{Neural computation}{18}{4}{904--941}.
\PrintBackRefs{\CurrentBib}

\bibitem [\protect \citeauthoryear {%
Leibold%
\ \BBA {} Kempter%
}{%
Leibold%
\ \BBA {} Kempter%
}{%
{\protect \APACyear {2008}}%
}]{%
lk08}
\APACinsertmetastar {%
lk08}%
\begin{APACrefauthors}%
Leibold, C.%
\BCBT {}\ \BBA {} Kempter, R.%
\end{APACrefauthors}%
\unskip\
\newblock
\APACrefYearMonthDay{2008}{Jan}{}.
\newblock
{\BBOQ}\APACrefatitle {{{S}parseness constrains the prolongation of memory
  lifetime via synaptic metaplasticity}} {{{S}parseness constrains the
  prolongation of memory lifetime via synaptic metaplasticity}}.{\BBCQ}
\newblock
\APACjournalVolNumPages{Cereb. Cortex}{18}{}{67--77}.
\PrintBackRefs{\CurrentBib}

\bibitem [\protect \citeauthoryear {%
Lillicrap%
, Cownden%
, Tweed%
\BCBL {}\ \BBA {} Akerman%
}{%
Lillicrap%
\ \protect \BOthers {.}}{%
{\protect \APACyear {2016}}%
}]{%
Lillicrap2016}
\APACinsertmetastar {%
Lillicrap2016}%
\begin{APACrefauthors}%
Lillicrap, T\BPBI P.%
, Cownden, D.%
, Tweed, D\BPBI B.%
\BCBL {}\ \BBA {} Akerman, C\BPBI J.%
\end{APACrefauthors}%
\unskip\
\newblock
\APACrefYearMonthDay{2016}{}{}.
\newblock
{\BBOQ}\APACrefatitle {Random synaptic feedback weights support error
  backpropagation for deep learning} {Random synaptic feedback weights support
  error backpropagation for deep learning}.{\BBCQ}
\newblock
\APACjournalVolNumPages{Nature Communications}{7}{}{}.
\PrintBackRefs{\CurrentBib}

\bibitem [\protect \citeauthoryear {%
J.~Lisman%
, Schulman%
\BCBL {}\ \BBA {} Cline%
}{%
J.~Lisman%
\ \protect \BOthers {.}}{%
{\protect \APACyear {2002}}%
}]{%
Lisman2002}
\APACinsertmetastar {%
Lisman2002}%
\begin{APACrefauthors}%
Lisman, J.%
, Schulman, H.%
\BCBL {}\ \BBA {} Cline, H.%
\end{APACrefauthors}%
\unskip\
\newblock
\APACrefYearMonthDay{2002}{}{}.
\newblock
{\BBOQ}\APACrefatitle {The molecular basis of CaMKII function in synaptic and
  behavioural memory} {The molecular basis of camkii function in synaptic and
  behavioural memory}.{\BBCQ}
\newblock
\APACjournalVolNumPages{Nature Reviews Neuroscience}{3}{3}{175--190}.
\PrintBackRefs{\CurrentBib}

\bibitem [\protect \citeauthoryear {%
J\BPBI E.~Lisman%
}{%
J\BPBI E.~Lisman%
}{%
{\protect \APACyear {1985}}%
}]{%
lisman1985}
\APACinsertmetastar {%
lisman1985}%
\begin{APACrefauthors}%
Lisman, J\BPBI E.%
\end{APACrefauthors}%
\unskip\
\newblock
\APACrefYearMonthDay{1985}{May}{}.
\newblock
{\BBOQ}\APACrefatitle {{{A} mechanism for memory storage insensitive to
  molecular turnover: a bistable autophosphorylating kinase}} {{{A} mechanism
  for memory storage insensitive to molecular turnover: a bistable
  autophosphorylating kinase}}.{\BBCQ}
\newblock
\APACjournalVolNumPages{Proc. Natl. Acad. Sci. U.S.A.}{82}{9}{3055--3057}.
\PrintBackRefs{\CurrentBib}

\bibitem [\protect \citeauthoryear {%
J\BPBI E.~Lisman%
\ \BBA {} Zhabotinsky%
}{%
J\BPBI E.~Lisman%
\ \BBA {} Zhabotinsky%
}{%
{\protect \APACyear {2001}}%
}]{%
lismanzhabotinsky2001}
\APACinsertmetastar {%
lismanzhabotinsky2001}%
\begin{APACrefauthors}%
Lisman, J\BPBI E.%
\BCBT {}\ \BBA {} Zhabotinsky, A\BPBI M.%
\end{APACrefauthors}%
\unskip\
\newblock
\APACrefYearMonthDay{2001}{Aug}{}.
\newblock
{\BBOQ}\APACrefatitle {{{A} model of synaptic memory: a
  {C}a{M}{K}{I}{I}/{P}{P}1 switch that potentiates transmission by organizing
  an {A}{M}{P}{A} receptor anchoring assembly}} {{{A} model of synaptic memory:
  a {C}a{M}{K}{I}{I}/{P}{P}1 switch that potentiates transmission by organizing
  an {A}{M}{P}{A} receptor anchoring assembly}}.{\BBCQ}
\newblock
\APACjournalVolNumPages{Neuron}{31}{}{191--201}.
\PrintBackRefs{\CurrentBib}

\bibitem [\protect \citeauthoryear {%
Marr%
}{%
Marr%
}{%
{\protect \APACyear {1971}}%
}]{%
marr71}
\APACinsertmetastar {%
marr71}%
\begin{APACrefauthors}%
Marr, D.%
\end{APACrefauthors}%
\unskip\
\newblock
\APACrefYearMonthDay{1971}{}{}.
\newblock
{\BBOQ}\APACrefatitle {Simple memory : {A} theory for archicortex} {Simple
  memory : {A} theory for archicortex}.{\BBCQ}
\newblock
\APACjournalVolNumPages{Philosophical Transactions of the Royal Society of
  London B}{262}{}{23-81}.
\PrintBackRefs{\CurrentBib}

\bibitem [\protect \citeauthoryear {%
McClelland%
\ \BBA {} Goddard%
}{%
McClelland%
\ \BBA {} Goddard%
}{%
{\protect \APACyear {1996}}%
}]{%
Mcclelland1996}
\APACinsertmetastar {%
Mcclelland1996}%
\begin{APACrefauthors}%
McClelland, J\BPBI L.%
\BCBT {}\ \BBA {} Goddard, N\BPBI H.%
\end{APACrefauthors}%
\unskip\
\newblock
\APACrefYearMonthDay{1996}{}{}.
\newblock
{\BBOQ}\APACrefatitle {Considerations arising from a complementary learning
  systems perspective on hippocampus and neocortex} {Considerations arising
  from a complementary learning systems perspective on hippocampus and
  neocortex}.{\BBCQ}
\newblock
\APACjournalVolNumPages{Hippocampus}{6}{6}{654--665}.
\PrintBackRefs{\CurrentBib}

\bibitem [\protect \citeauthoryear {%
McClelland%
, McNaughton%
\BCBL {}\ \BBA {} O'Reilly%
}{%
McClelland%
\ \protect \BOthers {.}}{%
{\protect \APACyear {1995}}%
}]{%
mcclelland95}
\APACinsertmetastar {%
mcclelland95}%
\begin{APACrefauthors}%
McClelland, J\BPBI L.%
, McNaughton, B\BPBI L.%
\BCBL {}\ \BBA {} O'Reilly, R\BPBI C.%
\end{APACrefauthors}%
\unskip\
\newblock
\APACrefYearMonthDay{1995}{}{}.
\newblock
{\BBOQ}\APACrefatitle {{{W}hy there are complementary learning systems in the
  hippocampus and neocortex: insights from the successes and failures of
  connectionist models of learning and memory}} {{{W}hy there are complementary
  learning systems in the hippocampus and neocortex: insights from the
  successes and failures of connectionist models of learning and
  memory}}.{\BBCQ}
\newblock
\APACjournalVolNumPages{Psychol Rev}{102}{}{419--457}.
\PrintBackRefs{\CurrentBib}

\bibitem [\protect \citeauthoryear {%
McCloskey%
\ \BBA {} Cohen%
}{%
McCloskey%
\ \BBA {} Cohen%
}{%
{\protect \APACyear {1989}}%
}]{%
McCloskey1989}
\APACinsertmetastar {%
McCloskey1989}%
\begin{APACrefauthors}%
McCloskey, M.%
\BCBT {}\ \BBA {} Cohen, N\BPBI J.%
\end{APACrefauthors}%
\unskip\
\newblock
\APACrefYearMonthDay{1989}{}{}.
\newblock
{\BBOQ}\APACrefatitle {Catastrophic interference in connectionist networks: The
  sequential learning problem} {Catastrophic interference in connectionist
  networks: The sequential learning problem}.{\BBCQ}
\newblock
\BIn{} \APACrefbtitle {Psychology of learning and motivation} {Psychology of
  learning and motivation}\ (\BVOL~24, \BPGS\ 109--165).
\newblock
\APACaddressPublisher{}{Elsevier}.
\PrintBackRefs{\CurrentBib}

\bibitem [\protect \citeauthoryear {%
McNaughton%
\ \BBA {} Morris%
}{%
McNaughton%
\ \BBA {} Morris%
}{%
{\protect \APACyear {1987}}%
}]{%
Mcnaughton1987}
\APACinsertmetastar {%
Mcnaughton1987}%
\begin{APACrefauthors}%
McNaughton, B\BPBI L.%
\BCBT {}\ \BBA {} Morris, R\BPBI G.%
\end{APACrefauthors}%
\unskip\
\newblock
\APACrefYearMonthDay{1987}{}{}.
\newblock
{\BBOQ}\APACrefatitle {Hippocampal synaptic enhancement and information storage
  within a distributed memory system} {Hippocampal synaptic enhancement and
  information storage within a distributed memory system}.{\BBCQ}
\newblock
\APACjournalVolNumPages{Trends in neurosciences}{10}{10}{408--415}.
\PrintBackRefs{\CurrentBib}

\bibitem [\protect \citeauthoryear {%
Mead%
}{%
Mead%
}{%
{\protect \APACyear {1990}}%
}]{%
mead}
\APACinsertmetastar {%
mead}%
\begin{APACrefauthors}%
Mead, C.%
\end{APACrefauthors}%
\unskip\
\newblock
\APACrefYearMonthDay{1990}{}{}.
\newblock
{\BBOQ}\APACrefatitle {Neuromorphic electronic systems} {Neuromorphic
  electronic systems}.{\BBCQ}
\newblock
\APACjournalVolNumPages{Proceedings of the IEEE}{78}{10}{1629--1636}.
\PrintBackRefs{\CurrentBib}

\bibitem [\protect \citeauthoryear {%
Miller%
, Zhabotinsky%
, Lisman%
\BCBL {}\ \BBA {} Wang%
}{%
Miller%
\ \protect \BOthers {.}}{%
{\protect \APACyear {2005}}%
}]{%
miller2005}
\APACinsertmetastar {%
miller2005}%
\begin{APACrefauthors}%
Miller, P.%
, Zhabotinsky, A\BPBI M.%
, Lisman, J\BPBI E.%
\BCBL {}\ \BBA {} Wang, X\BPBI J.%
\end{APACrefauthors}%
\unskip\
\newblock
\APACrefYearMonthDay{2005}{}{}.
\newblock
{\BBOQ}\APACrefatitle {{{T}he stability of a stochastic {C}a{M}{K}{I}{I}
  switch: dependence on the number of enzyme molecules and protein turnover}}
  {{{T}he stability of a stochastic {C}a{M}{K}{I}{I} switch: dependence on the
  number of enzyme molecules and protein turnover}}.{\BBCQ}
\newblock
\APACjournalVolNumPages{PLoS Biol.}{3}{}{e107}.
\PrintBackRefs{\CurrentBib}

\bibitem [\protect \citeauthoryear {%
Mongillo%
, Amit%
\BCBL {}\ \BBA {} Brunel%
}{%
Mongillo%
\ \protect \BOthers {.}}{%
{\protect \APACyear {2003}}%
}]{%
Mongillo2003}
\APACinsertmetastar {%
Mongillo2003}%
\begin{APACrefauthors}%
Mongillo, G.%
, Amit, D\BPBI J.%
\BCBL {}\ \BBA {} Brunel, N.%
\end{APACrefauthors}%
\unskip\
\newblock
\APACrefYearMonthDay{2003}{}{}.
\newblock
{\BBOQ}\APACrefatitle {Retrospective and prospective persistent activity
  induced by Hebbian learning in a recurrent cortical network} {Retrospective
  and prospective persistent activity induced by hebbian learning in a
  recurrent cortical network}.{\BBCQ}
\newblock
\APACjournalVolNumPages{European Journal of Neuroscience}{18}{7}{2011--2024}.
\PrintBackRefs{\CurrentBib}

\bibitem [\protect \citeauthoryear {%
Murray%
, Pye%
\BCBL {}\ \BBA {} Hockley%
}{%
Murray%
\ \protect \BOthers {.}}{%
{\protect \APACyear {1976}}%
}]{%
Murray1976}
\APACinsertmetastar {%
Murray1976}%
\begin{APACrefauthors}%
Murray, D.%
, Pye, C.%
\BCBL {}\ \BBA {} Hockley, W.%
\end{APACrefauthors}%
\unskip\
\newblock
\APACrefYearMonthDay{1976}{}{}.
\newblock
{\BBOQ}\APACrefatitle {Standing's power function in long-term memory}
  {Standing's power function in long-term memory}.{\BBCQ}
\newblock
\APACjournalVolNumPages{Psychological Research}{38}{4}{319--331}.
\PrintBackRefs{\CurrentBib}

\bibitem [\protect \citeauthoryear {%
Naim%
, Katkov%
, Romani%
\BCBL {}\ \BBA {} Tsodyks%
}{%
Naim%
\ \protect \BOthers {.}}{%
{\protect \APACyear {2019}}%
}]{%
Naim2019}
\APACinsertmetastar {%
Naim2019}%
\begin{APACrefauthors}%
Naim, M.%
, Katkov, M.%
, Romani, S.%
\BCBL {}\ \BBA {} Tsodyks, M.%
\end{APACrefauthors}%
\unskip\
\newblock
\APACrefYearMonthDay{2019}{}{}.
\newblock
{\BBOQ}\APACrefatitle {Fundamental Law of Memory Recall} {Fundamental law of
  memory recall}.{\BBCQ}
\newblock
\APACjournalVolNumPages{arXiv preprint arXiv:1905.02403}{}{}{}.
\PrintBackRefs{\CurrentBib}

\bibitem [\protect \citeauthoryear {%
O'Connor%
, Wittenberg%
\BCBL {}\ \BBA {} Wang%
}{%
O'Connor%
\ \protect \BOthers {.}}{%
{\protect \APACyear {2005}}%
}]{%
oconnor2005}
\APACinsertmetastar {%
oconnor2005}%
\begin{APACrefauthors}%
O'Connor, D\BPBI H.%
, Wittenberg, G\BPBI M.%
\BCBL {}\ \BBA {} Wang, S\BPBI S\BHBI H.%
\end{APACrefauthors}%
\unskip\
\newblock
\APACrefYearMonthDay{2005}{}{}.
\newblock
{\BBOQ}\APACrefatitle {Graded bidirectional synaptic plasticity is composed of
  switch-like unitary events} {Graded bidirectional synaptic plasticity is
  composed of switch-like unitary events}.{\BBCQ}
\newblock
\APACjournalVolNumPages{Proceedings of the National Academy of
  Sciences}{102}{27}{9679--9684}.
\PrintBackRefs{\CurrentBib}

\bibitem [\protect \citeauthoryear {%
Olshausen%
\ \BBA {} Field%
}{%
Olshausen%
\ \BBA {} Field%
}{%
{\protect \APACyear {2004}}%
}]{%
olshausen2004sparse}
\APACinsertmetastar {%
olshausen2004sparse}%
\begin{APACrefauthors}%
Olshausen, B\BPBI A.%
\BCBT {}\ \BBA {} Field, D\BPBI J.%
\end{APACrefauthors}%
\unskip\
\newblock
\APACrefYearMonthDay{2004}{}{}.
\newblock
{\BBOQ}\APACrefatitle {Sparse coding of sensory inputs} {Sparse coding of
  sensory inputs}.{\BBCQ}
\newblock
\APACjournalVolNumPages{Current opinion in neurobiology}{14}{4}{481--487}.
\PrintBackRefs{\CurrentBib}

\bibitem [\protect \citeauthoryear {%
O'Reilly%
\ \BBA {} Frank%
}{%
O'Reilly%
\ \BBA {} Frank%
}{%
{\protect \APACyear {2006}}%
}]{%
oreillyfrank2006}
\APACinsertmetastar {%
oreillyfrank2006}%
\begin{APACrefauthors}%
O'Reilly, R\BPBI C.%
\BCBT {}\ \BBA {} Frank, M\BPBI J.%
\end{APACrefauthors}%
\unskip\
\newblock
\APACrefYearMonthDay{2006}{Feb}{}.
\newblock
{\BBOQ}\APACrefatitle {Making working memory work: a computational model of
  learning in the prefrontal cortex and basal ganglia.} {Making working memory
  work: a computational model of learning in the prefrontal cortex and basal
  ganglia.}{\BBCQ}
\newblock
\APACjournalVolNumPages{Neural Comput}{18}{2}{283--328}.
\newblock
\begin{APACrefURL} \url{http://dx.doi.org/10.1162/089976606775093909}
  \end{APACrefURL}
\newblock
\begin{APACrefDOI} \doi{10.1162/089976606775093909} \end{APACrefDOI}
\PrintBackRefs{\CurrentBib}

\bibitem [\protect \citeauthoryear {%
O'Reilly%
\ \BBA {} McClelland%
}{%
O'Reilly%
\ \BBA {} McClelland%
}{%
{\protect \APACyear {1994}}%
}]{%
OReilly1994}
\APACinsertmetastar {%
OReilly1994}%
\begin{APACrefauthors}%
O'Reilly, R\BPBI C.%
\BCBT {}\ \BBA {} McClelland, J\BPBI L.%
\end{APACrefauthors}%
\unskip\
\newblock
\APACrefYearMonthDay{1994}{}{}.
\newblock
{\BBOQ}\APACrefatitle {Hippocampal conjunctive encoding, storage, and recall:
  avoiding a trade-off} {Hippocampal conjunctive encoding, storage, and recall:
  avoiding a trade-off}.{\BBCQ}
\newblock
\APACjournalVolNumPages{Hippocampus}{4}{6}{661--682}.
\PrintBackRefs{\CurrentBib}

\bibitem [\protect \citeauthoryear {%
Redondo%
\ \BBA {} Morris%
}{%
Redondo%
\ \BBA {} Morris%
}{%
{\protect \APACyear {2011}}%
}]{%
Redondo2011}
\APACinsertmetastar {%
Redondo2011}%
\begin{APACrefauthors}%
Redondo, R\BPBI L.%
\BCBT {}\ \BBA {} Morris, R\BPBI G.%
\end{APACrefauthors}%
\unskip\
\newblock
\APACrefYearMonthDay{2011}{}{}.
\newblock
{\BBOQ}\APACrefatitle {Making memories last: the synaptic tagging and capture
  hypothesis} {Making memories last: the synaptic tagging and capture
  hypothesis}.{\BBCQ}
\newblock
\APACjournalVolNumPages{Nature Reviews Neuroscience}{12}{1}{17--30}.
\PrintBackRefs{\CurrentBib}

\bibitem [\protect \citeauthoryear {%
Reymann%
\ \BBA {} Frey%
}{%
Reymann%
\ \BBA {} Frey%
}{%
{\protect \APACyear {2007}}%
}]{%
Reymann2007}
\APACinsertmetastar {%
Reymann2007}%
\begin{APACrefauthors}%
Reymann, K\BPBI G.%
\BCBT {}\ \BBA {} Frey, J\BPBI U.%
\end{APACrefauthors}%
\unskip\
\newblock
\APACrefYearMonthDay{2007}{}{}.
\newblock
{\BBOQ}\APACrefatitle {The late maintenance of hippocampal LTP: requirements,
  phases, synaptic tagging} {The late maintenance of hippocampal ltp:
  requirements, phases, synaptic tagging}.{\BBCQ}
\newblock
\APACjournalVolNumPages{Neuropharmacology}{52}{1}{24--40}.
\PrintBackRefs{\CurrentBib}

\bibitem [\protect \citeauthoryear {%
Richards%
\ \protect \BOthers {.}}{%
Richards%
\ \protect \BOthers {.}}{%
{\protect \APACyear {2019}}%
}]{%
Richards2019}
\APACinsertmetastar {%
Richards2019}%
\begin{APACrefauthors}%
Richards, B\BPBI A.%
, Lillicrap, T\BPBI P.%
, Beaudoin, P.%
, Bengio, Y.%
, Bogacz, R.%
, Christensen, A.%
\BDBL {}others%
\end{APACrefauthors}%
\unskip\
\newblock
\APACrefYearMonthDay{2019}{}{}.
\newblock
{\BBOQ}\APACrefatitle {A deep learning framework for neuroscience} {A deep
  learning framework for neuroscience}.{\BBCQ}
\newblock
\APACjournalVolNumPages{Nature neuroscience}{22}{11}{1761--1770}.
\PrintBackRefs{\CurrentBib}

\bibitem [\protect \citeauthoryear {%
Rigotti%
, Rubin%
, Wang%
\BCBL {}\ \BBA {} Fusi%
}{%
Rigotti%
\ \protect \BOthers {.}}{%
{\protect \APACyear {2010}}%
}]{%
rigotti10}
\APACinsertmetastar {%
rigotti10}%
\begin{APACrefauthors}%
Rigotti, M.%
, Rubin, D\BPBI B.%
, Wang, X\BHBI J.%
\BCBL {}\ \BBA {} Fusi, S.%
\end{APACrefauthors}%
\unskip\
\newblock
\APACrefYearMonthDay{2010}{}{}.
\newblock
{\BBOQ}\APACrefatitle {{{I}nternal representation of task rules by recurrent
  dynamics: the importance of the diversity of neural responses}} {{{I}nternal
  representation of task rules by recurrent dynamics: the importance of the
  diversity of neural responses}}.{\BBCQ}
\newblock
\APACjournalVolNumPages{Front Comput Neurosci}{4}{}{24}.
\PrintBackRefs{\CurrentBib}

\bibitem [\protect \citeauthoryear {%
Rizzuto%
\ \BBA {} Kahana%
}{%
Rizzuto%
\ \BBA {} Kahana%
}{%
{\protect \APACyear {2001}}%
}]{%
Rizzuto2001}
\APACinsertmetastar {%
Rizzuto2001}%
\begin{APACrefauthors}%
Rizzuto, D\BPBI S.%
\BCBT {}\ \BBA {} Kahana, M\BPBI J.%
\end{APACrefauthors}%
\unskip\
\newblock
\APACrefYearMonthDay{2001}{}{}.
\newblock
{\BBOQ}\APACrefatitle {An autoassociative neural network model of
  paired-associate learning} {An autoassociative neural network model of
  paired-associate learning}.{\BBCQ}
\newblock
\APACjournalVolNumPages{Neural Computation}{13}{9}{2075--2092}.
\PrintBackRefs{\CurrentBib}

\bibitem [\protect \citeauthoryear {%
Romani%
, Pinkoviezky%
, Rubin%
\BCBL {}\ \BBA {} Tsodyks%
}{%
Romani%
\ \protect \BOthers {.}}{%
{\protect \APACyear {2013}}%
}]{%
Romani2013}
\APACinsertmetastar {%
Romani2013}%
\begin{APACrefauthors}%
Romani, S.%
, Pinkoviezky, I.%
, Rubin, A.%
\BCBL {}\ \BBA {} Tsodyks, M.%
\end{APACrefauthors}%
\unskip\
\newblock
\APACrefYearMonthDay{2013}{}{}.
\newblock
{\BBOQ}\APACrefatitle {Scaling laws of associative memory retrieval} {Scaling
  laws of associative memory retrieval}.{\BBCQ}
\newblock
\APACjournalVolNumPages{Neural computation}{25}{10}{2523--2544}.
\PrintBackRefs{\CurrentBib}

\bibitem [\protect \citeauthoryear {%
Rosenblatt%
}{%
Rosenblatt%
}{%
{\protect \APACyear {1958}}%
}]{%
r58}
\APACinsertmetastar {%
r58}%
\begin{APACrefauthors}%
Rosenblatt, F.%
\end{APACrefauthors}%
\unskip\
\newblock
\APACrefYearMonthDay{1958}{}{}.
\newblock
{\BBOQ}\APACrefatitle {{The perceptron: a probabilistic model for information
  storage and organization in the brain}} {{The perceptron: a probabilistic
  model for information storage and organization in the brain}}.{\BBCQ}
\newblock
\APACjournalVolNumPages{Psychological Review}{65}{}{386-408}.
\newblock
\APACrefnote{Reprinted in: Anderson and Rosenfeld (eds.), {\em Neurocomputing:
  Foundations of Research}}
\PrintBackRefs{\CurrentBib}

\bibitem [\protect \citeauthoryear {%
Rosenblatt%
}{%
Rosenblatt%
}{%
{\protect \APACyear {1962}}%
}]{%
Rosenblatt62}
\APACinsertmetastar {%
Rosenblatt62}%
\begin{APACrefauthors}%
Rosenblatt, F.%
\end{APACrefauthors}%
\unskip\
\newblock
\APACrefYear{1962}.
\newblock
\APACrefbtitle {Principles of Neurodynamics} {Principles of neurodynamics}.
\newblock
\APACaddressPublisher{New York}{Spartan Books}.
\PrintBackRefs{\CurrentBib}

\bibitem [\protect \citeauthoryear {%
Roxin%
\ \BBA {} Fusi%
}{%
Roxin%
\ \BBA {} Fusi%
}{%
{\protect \APACyear {2013}}%
}]{%
rf13}
\APACinsertmetastar {%
rf13}%
\begin{APACrefauthors}%
Roxin, A.%
\BCBT {}\ \BBA {} Fusi, S.%
\end{APACrefauthors}%
\unskip\
\newblock
\APACrefYearMonthDay{2013}{Jul}{}.
\newblock
{\BBOQ}\APACrefatitle {Efficient partitioning of memory systems and its
  importance for memory consolidation.} {Efficient partitioning of memory
  systems and its importance for memory consolidation.}{\BBCQ}
\newblock
\APACjournalVolNumPages{PLoS Comput Biol}{9}{7}{e1003146}.
\newblock
\begin{APACrefURL} \url{http://dx.doi.org/10.1371/journal.pcbi.1003146}
  \end{APACrefURL}
\newblock
\begin{APACrefDOI} \doi{10.1371/journal.pcbi.1003146} \end{APACrefDOI}
\PrintBackRefs{\CurrentBib}

\bibitem [\protect \citeauthoryear {%
Rumelhart%
, Hinton%
\BCBL {}\ \BBA {} Williams%
}{%
Rumelhart%
\ \protect \BOthers {.}}{%
{\protect \APACyear {1986}}%
}]{%
rumelhart1986learning}
\APACinsertmetastar {%
rumelhart1986learning}%
\begin{APACrefauthors}%
Rumelhart, D\BPBI E.%
, Hinton, G\BPBI E.%
\BCBL {}\ \BBA {} Williams, R\BPBI J.%
\end{APACrefauthors}%
\unskip\
\newblock
\APACrefYearMonthDay{1986}{}{}.
\newblock
{\BBOQ}\APACrefatitle {Learning representations by back-propagating errors}
  {Learning representations by back-propagating errors}.{\BBCQ}
\newblock
\APACjournalVolNumPages{Nature}{323}{}{533--536}.
\PrintBackRefs{\CurrentBib}

\bibitem [\protect \citeauthoryear {%
Savin%
, Dayan%
\BCBL {}\ \BBA {} Lengyel%
}{%
Savin%
\ \protect \BOthers {.}}{%
{\protect \APACyear {2014}}%
}]{%
Savin2014}
\APACinsertmetastar {%
Savin2014}%
\begin{APACrefauthors}%
Savin, C.%
, Dayan, P.%
\BCBL {}\ \BBA {} Lengyel, M.%
\end{APACrefauthors}%
\unskip\
\newblock
\APACrefYearMonthDay{2014}{}{}.
\newblock
{\BBOQ}\APACrefatitle {Optimal recall from bounded metaplastic synapses:
  predicting functional adaptations in hippocampal area CA3} {Optimal recall
  from bounded metaplastic synapses: predicting functional adaptations in
  hippocampal area ca3}.{\BBCQ}
\newblock
\APACjournalVolNumPages{PLoS Comput Biol}{10}{2}{e1003489}.
\PrintBackRefs{\CurrentBib}

\bibitem [\protect \citeauthoryear {%
Scellier%
\ \BBA {} Bengio%
}{%
Scellier%
\ \BBA {} Bengio%
}{%
{\protect \APACyear {2016}}%
}]{%
Scellier2016}
\APACinsertmetastar {%
Scellier2016}%
\begin{APACrefauthors}%
Scellier, B.%
\BCBT {}\ \BBA {} Bengio, Y.%
\end{APACrefauthors}%
\unskip\
\newblock
\APACrefYearMonthDay{2016}{}{}.
\newblock
{\BBOQ}\APACrefatitle {Towards a biologically plausible backprop} {Towards a
  biologically plausible backprop}.{\BBCQ}
\newblock
\APACjournalVolNumPages{arXiv preprint arXiv:1602.05179}{}{}{}.
\PrintBackRefs{\CurrentBib}

\bibitem [\protect \citeauthoryear {%
Schapiro%
, Turk-Browne%
, Botvinick%
\BCBL {}\ \BBA {} Norman%
}{%
Schapiro%
\ \protect \BOthers {.}}{%
{\protect \APACyear {2017}}%
}]{%
schapiro2017}
\APACinsertmetastar {%
schapiro2017}%
\begin{APACrefauthors}%
Schapiro, A\BPBI C.%
, Turk-Browne, N\BPBI B.%
, Botvinick, M\BPBI M.%
\BCBL {}\ \BBA {} Norman, K\BPBI A.%
\end{APACrefauthors}%
\unskip\
\newblock
\APACrefYearMonthDay{2017}{}{}.
\newblock
{\BBOQ}\APACrefatitle {Complementary learning systems within the hippocampus: a
  neural network modelling approach to reconciling episodic memory with
  statistical learning} {Complementary learning systems within the hippocampus:
  a neural network modelling approach to reconciling episodic memory with
  statistical learning}.{\BBCQ}
\newblock
\APACjournalVolNumPages{Philosophical Transactions of the Royal Society B:
  Biological Sciences}{372}{1711}{20160049}.
\PrintBackRefs{\CurrentBib}

\bibitem [\protect \citeauthoryear {%
Sejnowski%
}{%
Sejnowski%
}{%
{\protect \APACyear {1977}}%
}]{%
sejnowski77}
\APACinsertmetastar {%
sejnowski77}%
\begin{APACrefauthors}%
Sejnowski, T\BPBI J.%
\end{APACrefauthors}%
\unskip\
\newblock
\APACrefYearMonthDay{1977}{}{}.
\newblock
{\BBOQ}\APACrefatitle {Storing covariance with nonlinearly interacting neurons}
  {Storing covariance with nonlinearly interacting neurons}.{\BBCQ}
\newblock
\APACjournalVolNumPages{J.~Math.~Biol.}{4}{}{303-}.
\PrintBackRefs{\CurrentBib}

\bibitem [\protect \citeauthoryear {%
Shouval%
}{%
Shouval%
}{%
{\protect \APACyear {2005}}%
}]{%
Shouval2005}
\APACinsertmetastar {%
Shouval2005}%
\begin{APACrefauthors}%
Shouval, H\BPBI Z.%
\end{APACrefauthors}%
\unskip\
\newblock
\APACrefYearMonthDay{2005}{}{}.
\newblock
{\BBOQ}\APACrefatitle {Clusters of interacting receptors can stabilize synaptic
  efficacies} {Clusters of interacting receptors can stabilize synaptic
  efficacies}.{\BBCQ}
\newblock
\APACjournalVolNumPages{Proceedings of the National Academy of Sciences of the
  United States of America}{102}{40}{14440--14445}.
\PrintBackRefs{\CurrentBib}

\bibitem [\protect \citeauthoryear {%
Shwartz-Ziv%
\ \BBA {} Tishby%
}{%
Shwartz-Ziv%
\ \BBA {} Tishby%
}{%
{\protect \APACyear {2017}}%
}]{%
Shwartz2017}
\APACinsertmetastar {%
Shwartz2017}%
\begin{APACrefauthors}%
Shwartz-Ziv, R.%
\BCBT {}\ \BBA {} Tishby, N.%
\end{APACrefauthors}%
\unskip\
\newblock
\APACrefYearMonthDay{2017}{}{}.
\newblock
{\BBOQ}\APACrefatitle {Opening the black box of deep neural networks via
  information} {Opening the black box of deep neural networks via
  information}.{\BBCQ}
\newblock
\APACjournalVolNumPages{arXiv preprint arXiv:1703.00810}{}{}{}.
\PrintBackRefs{\CurrentBib}

\bibitem [\protect \citeauthoryear {%
Squire%
\ \BBA {} Kandel%
}{%
Squire%
\ \BBA {} Kandel%
}{%
{\protect \APACyear {1999}}%
}]{%
squirekandel}
\APACinsertmetastar {%
squirekandel}%
\begin{APACrefauthors}%
Squire, L.%
\BCBT {}\ \BBA {} Kandel, E.%
\end{APACrefauthors}%
\unskip\
\newblock
\APACrefYear{1999}.
\newblock
\APACrefbtitle {Memory: from mind to molecules} {Memory: from mind to
  molecules}.
\newblock
\APACaddressPublisher{}{Scientific American Library}.
\PrintBackRefs{\CurrentBib}

\bibitem [\protect \citeauthoryear {%
Standing%
}{%
Standing%
}{%
{\protect \APACyear {1973}}%
}]{%
Standing1973}
\APACinsertmetastar {%
Standing1973}%
\begin{APACrefauthors}%
Standing, L.%
\end{APACrefauthors}%
\unskip\
\newblock
\APACrefYearMonthDay{1973}{}{}.
\newblock
{\BBOQ}\APACrefatitle {Learning 10000 pictures} {Learning 10000
  pictures}.{\BBCQ}
\newblock
\APACjournalVolNumPages{The Quarterly journal of experimental
  psychology}{25}{2}{207--222}.
\PrintBackRefs{\CurrentBib}

\bibitem [\protect \citeauthoryear {%
Treves%
}{%
Treves%
}{%
{\protect \APACyear {1990}}%
}]{%
Treves1990}
\APACinsertmetastar {%
Treves1990}%
\begin{APACrefauthors}%
Treves, A.%
\end{APACrefauthors}%
\unskip\
\newblock
\APACrefYearMonthDay{1990}{}{}.
\newblock
{\BBOQ}\APACrefatitle {Graded-response neurons and information encodings in
  autoassociative memories} {Graded-response neurons and information encodings
  in autoassociative memories}.{\BBCQ}
\newblock
\APACjournalVolNumPages{Physical Review A}{42}{4}{2418}.
\PrintBackRefs{\CurrentBib}

\bibitem [\protect \citeauthoryear {%
Treves%
\ \BBA {} Rolls%
}{%
Treves%
\ \BBA {} Rolls%
}{%
{\protect \APACyear {1991}}%
}]{%
tr91}
\APACinsertmetastar {%
tr91}%
\begin{APACrefauthors}%
Treves, A.%
\BCBT {}\ \BBA {} Rolls, E\BPBI T.%
\end{APACrefauthors}%
\unskip\
\newblock
\APACrefYearMonthDay{1991}{}{}.
\newblock
{\BBOQ}\APACrefatitle {What determines the capacity of autoassociative memories
  in the brain?} {What determines the capacity of autoassociative memories in
  the brain?}{\BBCQ}
\newblock
\APACjournalVolNumPages{Network: Computation in Neural
  Systems}{2}{4}{371--397}.
\PrintBackRefs{\CurrentBib}

\bibitem [\protect \citeauthoryear {%
Treves%
\ \BBA {} Rolls%
}{%
Treves%
\ \BBA {} Rolls%
}{%
{\protect \APACyear {1994}}%
}]{%
Treves1994}
\APACinsertmetastar {%
Treves1994}%
\begin{APACrefauthors}%
Treves, A.%
\BCBT {}\ \BBA {} Rolls, E\BPBI T.%
\end{APACrefauthors}%
\unskip\
\newblock
\APACrefYearMonthDay{1994}{}{}.
\newblock
{\BBOQ}\APACrefatitle {Computational analysis of the role of the hippocampus in
  memory} {Computational analysis of the role of the hippocampus in
  memory}.{\BBCQ}
\newblock
\APACjournalVolNumPages{Hippocampus}{4}{3}{374--391}.
\PrintBackRefs{\CurrentBib}

\bibitem [\protect \citeauthoryear {%
Tsodyks%
}{%
Tsodyks%
}{%
{\protect \APACyear {1990}}%
}]{%
t90}
\APACinsertmetastar {%
t90}%
\begin{APACrefauthors}%
Tsodyks, M.%
\end{APACrefauthors}%
\unskip\
\newblock
\APACrefYearMonthDay{1990}{}{}.
\newblock
{\BBOQ}\APACrefatitle {Associative memory in neural networks with binary
  synapses} {Associative memory in neural networks with binary
  synapses}.{\BBCQ}
\newblock
\APACjournalVolNumPages{Mod. Phys. Lett.}{B4}{}{713-716}.
\PrintBackRefs{\CurrentBib}

\bibitem [\protect \citeauthoryear {%
Tsodyks%
\ \BBA {} Feigel'man%
}{%
Tsodyks%
\ \BBA {} Feigel'man%
}{%
{\protect \APACyear {1988}}%
}]{%
tf88}
\APACinsertmetastar {%
tf88}%
\begin{APACrefauthors}%
Tsodyks, M.%
\BCBT {}\ \BBA {} Feigel'man, M\BPBI V.%
\end{APACrefauthors}%
\unskip\
\newblock
\APACrefYearMonthDay{1988}{}{}.
\newblock
{\BBOQ}\APACrefatitle {The enhanced storage capacity in neural networks with
  low activity level} {The enhanced storage capacity in neural networks with
  low activity level}.{\BBCQ}
\newblock
\APACjournalVolNumPages{Europhys.~Lett.}{46}{}{101-}.
\PrintBackRefs{\CurrentBib}

\bibitem [\protect \citeauthoryear {%
van~de Ven%
, Siegelmann%
\BCBL {}\ \BBA {} Tolias%
}{%
van~de Ven%
\ \protect \BOthers {.}}{%
{\protect \APACyear {2020}}%
}]{%
Van2020}
\APACinsertmetastar {%
Van2020}%
\begin{APACrefauthors}%
van~de Ven, G\BPBI M.%
, Siegelmann, H\BPBI T.%
\BCBL {}\ \BBA {} Tolias, A\BPBI S.%
\end{APACrefauthors}%
\unskip\
\newblock
\APACrefYearMonthDay{2020}{}{}.
\newblock
{\BBOQ}\APACrefatitle {Brain-inspired replay for continual learning with
  artificial neural networks} {Brain-inspired replay for continual learning
  with artificial neural networks}.{\BBCQ}
\newblock
\APACjournalVolNumPages{Nature communications}{11}{1}{1--14}.
\PrintBackRefs{\CurrentBib}

\bibitem [\protect \citeauthoryear {%
Van~Rossum%
, Shippi%
\BCBL {}\ \BBA {} Barrett%
}{%
Van~Rossum%
\ \protect \BOthers {.}}{%
{\protect \APACyear {2012}}%
}]{%
van2012soft}
\APACinsertmetastar {%
van2012soft}%
\begin{APACrefauthors}%
Van~Rossum, M\BPBI C.%
, Shippi, M.%
\BCBL {}\ \BBA {} Barrett, A\BPBI B.%
\end{APACrefauthors}%
\unskip\
\newblock
\APACrefYearMonthDay{2012}{}{}.
\newblock
{\BBOQ}\APACrefatitle {Soft-bound synaptic plasticity increases storage
  capacity} {Soft-bound synaptic plasticity increases storage capacity}.{\BBCQ}
\newblock
\APACjournalVolNumPages{PLoS Comput Biol}{8}{12}{e1002836}.
\PrintBackRefs{\CurrentBib}

\bibitem [\protect \citeauthoryear {%
Whittington%
\ \protect \BOthers {.}}{%
Whittington%
\ \protect \BOthers {.}}{%
{\protect \APACyear {2020}}%
}]{%
whittington2020}
\APACinsertmetastar {%
whittington2020}%
\begin{APACrefauthors}%
Whittington, J\BPBI C.%
, Muller, T\BPBI H.%
, Mark, S.%
, Chen, G.%
, Barry, C.%
, Burgess, N.%
\BCBL {}\ \BBA {} Behrens, T\BPBI E.%
\end{APACrefauthors}%
\unskip\
\newblock
\APACrefYearMonthDay{2020}{}{}.
\newblock
{\BBOQ}\APACrefatitle {The Tolman-Eichenbaum machine: Unifying space and
  relational memory through generalization in the hippocampal formation} {The
  tolman-eichenbaum machine: Unifying space and relational memory through
  generalization in the hippocampal formation}.{\BBCQ}
\newblock
\APACjournalVolNumPages{Cell}{183}{5}{1249--1263}.
\PrintBackRefs{\CurrentBib}

\bibitem [\protect \citeauthoryear {%
Willshaw%
, Buneman%
\BCBL {}\ \BBA {} Longuet-Higgins%
}{%
Willshaw%
\ \protect \BOthers {.}}{%
{\protect \APACyear {1969}}%
}]{%
w69}
\APACinsertmetastar {%
w69}%
\begin{APACrefauthors}%
Willshaw, D\BPBI J.%
, Buneman, O\BPBI P.%
\BCBL {}\ \BBA {} Longuet-Higgins, H\BPBI C.%
\end{APACrefauthors}%
\unskip\
\newblock
\APACrefYearMonthDay{1969}{}{}.
\newblock
{\BBOQ}\APACrefatitle {Non-holographic associative memory.} {Non-holographic
  associative memory.}{\BBCQ}
\newblock
\APACjournalVolNumPages{Nature}{}{}{}.
\PrintBackRefs{\CurrentBib}

\bibitem [\protect \citeauthoryear {%
Wixted%
\ \BBA {} Ebbesen%
}{%
Wixted%
\ \BBA {} Ebbesen%
}{%
{\protect \APACyear {1997}}%
}]{%
wixted1997}
\APACinsertmetastar {%
wixted1997}%
\begin{APACrefauthors}%
Wixted, J\BPBI T.%
\BCBT {}\ \BBA {} Ebbesen, E\BPBI B.%
\end{APACrefauthors}%
\unskip\
\newblock
\APACrefYearMonthDay{1997}{Sep}{}.
\newblock
{\BBOQ}\APACrefatitle {{{G}enuine power curves in forgetting: a quantitative
  analysis of individual subject forgetting functions}} {{{G}enuine power
  curves in forgetting: a quantitative analysis of individual subject
  forgetting functions}}.{\BBCQ}
\newblock
\APACjournalVolNumPages{Mem Cognit}{25}{}{731--739}.
\PrintBackRefs{\CurrentBib}

\bibitem [\protect \citeauthoryear {%
X.~Wu%
, Mel%
, Strouse%
\BCBL {}\ \BBA {} Mel%
}{%
X.~Wu%
\ \protect \BOthers {.}}{%
{\protect \APACyear {2019}}%
}]{%
Wu2019}
\APACinsertmetastar {%
Wu2019}%
\begin{APACrefauthors}%
Wu, X.%
, Mel, G\BPBI C.%
, Strouse, D.%
\BCBL {}\ \BBA {} Mel, B\BPBI W.%
\end{APACrefauthors}%
\unskip\
\newblock
\APACrefYearMonthDay{2019}{}{}.
\newblock
{\BBOQ}\APACrefatitle {How dendrites affect online recognition memory} {How
  dendrites affect online recognition memory}.{\BBCQ}
\newblock
\APACjournalVolNumPages{PLoS computational biology}{15}{5}{e1006892}.
\PrintBackRefs{\CurrentBib}

\bibitem [\protect \citeauthoryear {%
X\BPBI E.~Wu%
\ \BBA {} Mel%
}{%
X\BPBI E.~Wu%
\ \BBA {} Mel%
}{%
{\protect \APACyear {2009}}%
}]{%
wm09}
\APACinsertmetastar {%
wm09}%
\begin{APACrefauthors}%
Wu, X\BPBI E.%
\BCBT {}\ \BBA {} Mel, B\BPBI W.%
\end{APACrefauthors}%
\unskip\
\newblock
\APACrefYearMonthDay{2009}{}{}.
\newblock
{\BBOQ}\APACrefatitle {Capacity-enhancing synaptic learning rules in a medial
  temporal lobe online learning model} {Capacity-enhancing synaptic learning
  rules in a medial temporal lobe online learning model}.{\BBCQ}
\newblock
\APACjournalVolNumPages{Neuron}{62}{1}{31--41}.
\PrintBackRefs{\CurrentBib}

\bibitem [\protect \citeauthoryear {%
Yamins%
\ \protect \BOthers {.}}{%
Yamins%
\ \protect \BOthers {.}}{%
{\protect \APACyear {2014}}%
}]{%
Yamins2014}
\APACinsertmetastar {%
Yamins2014}%
\begin{APACrefauthors}%
Yamins, D\BPBI L.%
, Hong, H.%
, Cadieu, C\BPBI F.%
, Solomon, E\BPBI A.%
, Seibert, D.%
\BCBL {}\ \BBA {} DiCarlo, J\BPBI J.%
\end{APACrefauthors}%
\unskip\
\newblock
\APACrefYearMonthDay{2014}{}{}.
\newblock
{\BBOQ}\APACrefatitle {Performance-optimized hierarchical models predict neural
  responses in higher visual cortex} {Performance-optimized hierarchical models
  predict neural responses in higher visual cortex}.{\BBCQ}
\newblock
\APACjournalVolNumPages{Proceedings of the national academy of
  sciences}{111}{23}{8619--8624}.
\PrintBackRefs{\CurrentBib}

\bibitem [\protect \citeauthoryear {%
Zenke%
, Poole%
\BCBL {}\ \BBA {} Ganguli%
}{%
Zenke%
\ \protect \BOthers {.}}{%
{\protect \APACyear {2017}}%
}]{%
zenke2017}
\APACinsertmetastar {%
zenke2017}%
\begin{APACrefauthors}%
Zenke, F.%
, Poole, B.%
\BCBL {}\ \BBA {} Ganguli, S.%
\end{APACrefauthors}%
\unskip\
\newblock
\APACrefYearMonthDay{2017}{}{}.
\newblock
{\BBOQ}\APACrefatitle {Continual learning through synaptic intelligence}
  {Continual learning through synaptic intelligence}.{\BBCQ}
\newblock
\BIn{} \APACrefbtitle {Proceedings of the 34th International Conference on
  Machine Learning-Volume 70} {Proceedings of the 34th international conference
  on machine learning-volume 70}\ (\BPGS\ 3987--3995).
\PrintBackRefs{\CurrentBib}

\bibitem [\protect \citeauthoryear {%
Ziegler%
, Zenke%
, Kastner%
\BCBL {}\ \BBA {} Gerstner%
}{%
Ziegler%
\ \protect \BOthers {.}}{%
{\protect \APACyear {2015}}%
}]{%
Ziegler2015}
\APACinsertmetastar {%
Ziegler2015}%
\begin{APACrefauthors}%
Ziegler, L.%
, Zenke, F.%
, Kastner, D\BPBI B.%
\BCBL {}\ \BBA {} Gerstner, W.%
\end{APACrefauthors}%
\unskip\
\newblock
\APACrefYearMonthDay{2015}{}{}.
\newblock
{\BBOQ}\APACrefatitle {Synaptic consolidation: from synapses to behavioral
  modeling} {Synaptic consolidation: from synapses to behavioral
  modeling}.{\BBCQ}
\newblock
\APACjournalVolNumPages{The Journal of Neuroscience}{35}{3}{1319--1334}.
\PrintBackRefs{\CurrentBib}

\end{thebibliography}

\end{document}